\magnification=1200
\def \G {{\cal G}}
\def \p {\partial}
\def \t {\tilde}
\def \a  {\alpha}
\def \d  {\delta}
\def \L {\Lambda}
\def \drightarrow {\buildrel{\der}\over\longrightarrow}
\def \der {d_{_{E.L}}}
\def \C   {{\cal C}}
\def \B   {\overline}
\def \V {{\cal V}}
\def \vsubs {\cap\phantom{\vrule height2mm depth0mm width0.4pt}
                  \vrule height2mm depth0mm width0.4pt}
\def \U {{\cal U}}
\def \P {{\cal P}}
\def \Ga {\Gamma}
\def \min {{\cal n}}
\def \R  {{\bf R}}
\def \vare {\varepsilon}
\def \Hs {H^1(\G,\L^0(M))}
\def\Fad {1}
\def\Jack{2}
\def\Kost {3}
\def \Dub {4}
\def\Mcmull{5}
\def\Hen{6}
\def\Stash{7}
\def\Dix{8}
\def\Barn{9}
\def\Pig{10}
\def\Arn{11}
\def\Ann{12}
\def\Vor{13}
\def\Kh{14}
\def\Voron{15}
\def\Olv{16}
\def\Khud{17}
\def\Sch{18}
\def\Stern{19}
\def\Ners{20}
\def\Stor{21}
\def\Vin{22}
\def\Postn{23}
\def\Sta{24}

                               \hfill{physics/9712052}
                   \bigskip
 \centerline{\bf Double Complexes and Cohomological Hierarchy}
 \centerline {\bf in a Space of Weakly Invariant
           Lagrangians of Mechanics}
                   \bigskip

   \centerline   {\bf O.M. Khudaverdian}
  \centerline   {\it  Laboratory of Computing Technique and Automation,
                   Joint Institute for Nuclear Research}
    \centerline {\it Dubna, Moscow Region  141980, Russia,}
   \centerline {\it on leave of absence from
   Department of Theoretical Physics }
   \centerline
  {\it  of Yerevan State  University, 375049  Yerevan, Armenia}

 \centerline  {e-mails: "khudian@vxjinr.jinr.ru",
                          "khudian@sc2a.unige.ch"}

    \centerline  {\bf D.A. Sahakyan}
 \centerline {\it Department of Theoretical Physics, Yerevan State
                                           University}
 \centerline {\it and}
  \centerline   {\it  University Centre at
                   Joint Institute for Nuclear Research}
 \centerline  {e-mail: "sahakian@uniphi.yerphi.am}
\bigskip

 {\it For a given configuration space $M$ and Lie algebra
 ${\G}$ whose action is defined on $M$ the
 space $\V_{0.0}$ of weakly ${\G}$-invariant Lagrangians
 (i.e. Lagrangians
 whose motion equations left hand sides
 are ${\G}$-invariant) is studied.
 The problem is reformulated in terms of
   the double complex of Lie algebra cochains with
  values in the complex of
 Lagrangians.  Calculating the cohomology of this
 complex using the method of spectral sequences
 we arrive at the hierarchy in the space $\V_{0.0}$:
 The double filtration
 $\{\V_{s.\sigma}\}$ ($s=0,1,2,3,4,\sigma=0,1$)
 and the homomorphisms on every space $\V_{s.\sigma}$
are constructed.
 These homomorphisms take values in the cohomologies
 of the algebra $\G$ and configuration space $M$.
  On one hand every space $\V_{s,\sigma}$ is the kernel
 of the corresponding homomorphism, on the other hand
 this space is defined by its physical properties.}

\bigskip
            \centerline{\bf I Introduction}
   \medskip
 The cohomology of the symmetries algebra  has important
 consequences for properties of corresponding theory [\Fad,\Jack]
 and cohomological methods play essential
 role in many problems of modern
 fields theory.
 For example their application made more clear the
 understanding of algebraic origin of gauge anomalies.
 As it was shown in [\Fad] one can consider axial anomalies of
 four-dimensional gauge theory in terms of infinitesimal cocycles in a
 representation of gauge group.

 Another example is BRST formalism
which at beginning  was formulated
 in terms of symplectic geometry
 of phase space expanded by the ghosts and antighosts,
then it was understood
 [\Kost,\Dub,\Mcmull,\Hen,\Stash] that the language
of homological algebra is more
deeply related with
 physical meaning of this formalism: Inclusion of ghosts and
 antighosts corresponds to the construction of the chain of free modules
 (free resolvent) on phase space of constrained system where the
 constrains cannot be resolved in a direct way. The operator
 corresponding to BRST charge becomes the differential of the complex of
 these resolvents. Further the investigation
 of local BRST cohomology was performed with use of
 developed homological methods.
 (See [\Dix,\Barn,\Pig] with citations there.)

 In this paper we consider more modest problem.
 We study relations between
 Noether identities and related phenomena
 for global symmetries of Lagrangians and
 cohomological properties of the algebra of these symmetries.

 Our considerations will be carried out for  mechanics
  but the scheme  has  straightforward
 generalization on the case of field theory Lagrangians.

 The standard statement (Noether 1-st Theorem) is that if the Lagrangian
 $L$ is invariant under the action of Lie algebra
 ${\cal G}$ of rigid symmetries $\{\d_i\}$
 then to every
 symmetry $\d_i$ corresponds
 the charge $N_i(L)$ which is preserved on the
 equations of motion [\Arn].

If to $\{\d_i\}$ corresponds the Lie algebra of vector fields
$\{{\bf X}_i=X_i^\mu{\p\over\p q^\mu}\}$
 (infinitesimal transformations of configuration
 space)  then $\d_iq^\mu\sim X^\mu_i$,

 \line{\hss $\d_i L=0=
       {\cal L}_{{\bf X}_i}L=X\sb i\sp\mu
     {\cal F}_\mu(L)+{d\over dt}\left({X_i}\sp \a{\p L\over \p
    {\dot q}\sp \a}\right)$ and\hss}
                      $$
           N_i={X_i}\sp \mu{\p L\over \p {\dot q}\sp \mu},
        \qquad{\rm where}\qquad\qquad
        {\cal F}_\mu(L)={\p L\over \p q\sp \mu}-
        {d\over dt}{\p L\over \p {\dot q}\sp \mu}
                                         \eqno (1.1)
                     $$
is the  left hand side (l.h.s.) of the equations of motion
 ${\cal F}_\mu=0$ of the Lagrangian $L$.

The statement of Noether theorem is valid also
 in a case if under the actions of transformations
 $\{\d_i\}$ Lagrangian
is preserved up to a full derivative of some functions
 $\{\a_i(q)\}$
                              $$
    \d_i L=0 \rightarrow \d_i L=d\a_i,\,\,{\rm then}\,\,
   N_i(L)\rightarrow N_i(L)-{\a_i}\,.
                                       \eqno(1.2)
                             $$
 At what extent this full derivative is essential?
 The redefinition of $L$ on a full derivative
$L \rightarrow L+df$ changes
 $\a_i$ on $\a_i+\d_i f$.
 The algebra of symmetries of the Lagrangian can be considered
as generalized if
 $d\a_i$ is not equal $0$ in (1.2), and it is
 {\it essentially generalized} if
it cannot be canceled by redefinition
 of Lagrangian on a full derivative i.e.
 $\d_i L=d\a_i$ but the equations

  \line{\hss$ d(\a_i+\d_i f)=0\,.$\hss     (1.3)}
 have no solutions.

 Using the basic properties of
 operators $\d$ and $d$:
  $\d^2=d^2=0,\,d\d=\d d$
 (see  the Section II) we obtain from
    (1.2) that
                           $$
 0={\d}\sp 2 L=\d d\a_i=d\d\a_i,
     \,\,{\rm so}\,\, (\d\a)_{ij}=w_{ij}=constant,
                                       \eqno(1.4)
                           $$
      where
   $(\d\a)_{ij}={{\cal L}_i}\a_j-{{\cal L}_j}\a_i-
   \a_k c\sp k_{ij}$ and $c^k_{ij}$
 are structure constants of the symmetries Lie algebra.

 It is easy to see that $w_{ij}$ is the cocycle of algebra
 $\cal G$ in constants. In a case if $w_{ij}$ is not coboundary
 one can see that the symmetries are essentially generalized.
 Indeed if according to (1.3) $\a_i=-\d_i f+t_i$
where $t_i$ are constants then $w_{ij}$ in (1.4)
  is coboundary in constants:
 $w_{ij}=(\d t)_{ij}=-c_{ij}^k t_k$.

  Let us consider for example the algebra of space translations.
 This algebra has 2-cohomology  in constants which are represented
 by antisymmetric tensors $B_{ik}$. (This algebra is abelian, so
 $\d B=0$ and the equation $B=\d t$ has no solutions in constants.)
  To obtain Lagrangian
 which possesses generalized translation symmetries
 corresponding to these cocycles, we note that
   for this Lagrangian $\a_i=A_{ij} q\sp j$.
  By redefinition of a Lagrangian on a full
 derivative one can reduce $A_{ij}$ to antisymmetric tensor and we come
 to Lagrangian

\line{\hss$ L=f(\dot q)+q\sp i B_{ij} {\dot q}\sp j\,.
                                                 $\hss(1.5)}
 If $f(\dot q)={m {\dot q}\sp 2\over 2}$ it is the well-known
 Lagrangian of particle in constant magnetic field.

 In the Section $V$ we consider an analogous statement
 for Galilean group: we show that one comes to
 the Lagrangian of free
 particle as a unique Lagrangian corresponding
 to Bargman cocycle of Lie algebra of Galilean group.

 We see that one of the reasons of generalized symmetries
 appearing is the existence of 2-cohomology of corresponding
 Lie algebra [\Jack,\Ann].
 Of course situation is more complicated.
 For example by evident reasons for this
 phenomenon is responsible de Rham cohomology of configuration space.
   If $L_{inv}$ is $\G$-invariant Lagrangian and
     $L=L_{inv}+A_\mu(q){\dot q}^\mu$ where
 $A_\mu(q)dq^\mu$ is a closed differential $1$-form which is not exact
 ($A_\mu(q)dq^\mu\not=df$) then in general $L$ is not $\G$-invariant.
 It has the same equations of motion but it differs from
 $L_{inv}$  on
 Aharonov-Bohm like effects.

 Even in the case if de Rham cohomology is trivial and
 the cocycle $w_{ij}$ in (1.4) is coboundary
 the symmetries of Lagrangian
 can be essentially generalized. The coboundary condition
  $w_{ij}=-c^k_{ij}t_k$  is necessary but not sufficient for
 (1.3) to have a solution.
It is another cohomologies of  symmetries algebra
 which prevent a Lagrangian
 to be reduced to $\G$-invariant by redefinition on a
 full derivative.

 The purpose of our paper is to investigate
 systematically this phenomenon.

 For the algebra $\G$ of vector fields
  on the configuration
  space $M$
  and a Lagrangian $L(q,\dot q)$ on $M$
 we  considered
  the following possible cases of
 generalized symmetries appearing
                  $$
               \matrix
                   {
                   \hbox
 {1) The action of $\G$ on the Lagrangian $L$
                produces the $2$-cocycle on $\G$:}\cr
     \d_i L(q,\dot q)={d\over dt}\a_i(q),\,
              w_{ij}=\d_i\a_j-\d_j\a_i -c_{ij}^k\a_k\,.\cr
                      }
                     $$
                  $$
               \matrix
                   {
                 \hbox
       {2) The action of $\G$ on the Lagrangian $L$
             produces the $2$-cocycle, but it is trivial:}\cr
             w_{ij}=-c_{ij}^k t_k\,. \cr}
                         $$
                         $$
                      \matrix
                         {
                      \hbox
     {3) The Lagrangian $L$ differs from invariant one
                   on a closed form:}\cr
           L=L_{inv}+A_\mu(q)\dot q^\mu,
         (\p_\mu A_\nu-\p_\nu A_\mu=0)\cr
                  \hbox
   {hence
      $\d_i L={d\over dt}(A_\mu X^\mu_i)$
                  and $w_{ij}=0$.} \cr}
                         $$
                         $$
                      \matrix
                         {
                       \hbox
     {4) The Lagrangian $L$ differs from
               $\G$-invariant one
                   on an exact form (full derivative):}\cr
           L=L_{inv}+\p_\mu f(q)\dot q^\mu=
      L_{inv}+{d\over dt}f(q),\,\,\d_i L_{inv}=0.\cr}
                                            \eqno (1.6)
                      $$
 One can see that

 \line{\hss$"4"\Rightarrow "3"\Rightarrow
      "2"\Rightarrow "1"$\hss               (1.7)}

  We briefly discuss how the generalized symmetries reveal
 itself in Hamiltonian mechanics and in
 a quasiclassical approximation of quantum mechanics.

  If the Lagrangian is $\G$-invariant,
 then to the Noether charges $N_i(L)$ in (1.1)
 in Hamiltonian framework correspond the charges
 $N^{ham}_i=X_i^\mu p_\mu$. They generate the $\G$-algebra
 structure
 via Poisson brackets

 \line{\hss$\{N_i^{ham},N_j^{ham}\}=c_{ij}^k N_k^{ham}\,.$
                                   \hss (1.8)}

 In quasiclassical approximation of quantum mechanics
 to these charges correspond the operators
 $X_i^\mu {\hat p}_\mu$ . Their action on quasiclassical
 wave function in configuration representation
 is reduced to infinitesimal transformation of wave functions
 argument:

\noindent
\line{\hss
  $\imath{\hat\d}_i \Psi=\Psi(q^\mu+\d_i q^\mu)-\Psi(q^\mu)$.\hss}
 In the case if symmetries algebra is generalized,
 one can see that correspondingly to (1.2)

\line{\hss $N^{ham}_i=X_i^\mu p_\mu-\a_i.$\hss}
 The  corresponding operators act not only on quasiclassical
 wave
 functions argument but on its phase
 too:
            $$
    {\hat\d}_i \Psi=
     -\imath X^\mu_i{\p\Psi(q)\over \p q^\mu}+
                   \imath\a_i(q)\Psi(q^\mu)\,.
                                              \eqno (1.9)
             $$
 In the case if the Lagrangian does not possess the property
 $"2"$  in (1.6), i.e. the generalized symmetries
 lead to non-trivial cocycle,
the Lie algebra of Hamiltonian Noether
charges $N_i^{ham}$ is the central extension
of  the Lie algebra $\G$ which corresponds to
 the cohomology class of the cocycle $w_{ij}$.
                      $$
   \{N_i^{ham},N_j^{ham}\}=c_{ij}^k N_k^{ham}+w_{ij}\,.
                       \eqno (1.10)
                     $$

 Correspondingly in this case  in (1.9) is realized
 an essentially projective representation of the
 Lie algebra $\G$.

 In the case if the Lagrangian possesses the property
 $"2"$  in (1.6), then one can choose $\a_i$
 such that  (1.8) is satisfied and the quantum
  representation (1.9) of $\G$ becomes linear.
  But if this Lagrangian does not possess the property "4" in (1.6)
 then the action of quantum transformation on the phase factor
 cannot be removed by redefinition
   $\Psi\rightarrow e^{if}\Psi$ of the wave function corresponding to
 redefinition of Lagrangian on a full derivative.
  In this case one can say that the linear transformation (1.9)
   is splitted on a space-like transformation$+$intrinsic
  spin-like transformation.
  Nevertheless if the Lagrangian
  possesses the property "3", i.e. it differs
  of an invariant Lagrangian on Bohm-Aharonov like effects, then
  the action on  phase in (1.9) can be removed locally.

 We call time-independent Lagrangian $L(q,{\dot q})$
 {\it weakly $\G$-invariant} if l.h.s. of its
 motion equations (1.1) is $\G$-invariant.
 For example  the Lagrangian $L$ in (1.2) is weakly
 $\G$-invariant. One can show that if $L$ is weakly
 $\G$-invariant Lagrangian   then

   \line{\hss$\d_i L=c_i+w_i,$\hss             (1.11)}
 where $c_i$ are constants and
 $w_i$ correspond to closed forms:
 $w_i=w_{i\mu}(q){\dot q}$ where differential $1$-forms
  $w_{i\mu}(q)dq^\mu$ are closed.
 (See in details later.)

  If $\{w_i\}$ correspond to exact forms:
   $w_{i\mu}(q)dq^\mu=da_i(q)$,
  $w_{i\mu}(q){\dot q}^\mu
 =\p_\mu\a_i(q){\dot q}^\mu=d\a_i(q)/dt$
and

  \line{\hss $c_i=0$ \hss                          (1.12)}
  then we come to (1.2). In the case if (1.12)
is not obeyed the corresponding Noether charges
                         $$
       N_i={X_i}\sp\mu{\p L\over \p {\dot q}\sp \mu}-
                \a_i-c_it
                                             \eqno (1.13)
                         $$
 depend on time.

 We denote by  $\V_{0.0}$ the space of
 weakly $\G$-invariant Lagrangians
on $M$ and by $\V_{0.1}$ the subspace of $\V_{0.0}$
 for which the condition
 (1.12) is satisfied. We denote by
  $\V_{s.1}$ ($s=1,2,3,4,$) the space of Lagrangians for which
the property $"s"$ in (1.6) is satisfied.
 According to (1.7)
                           $$
 \V_{4.1}\subseteq \V_{3.1}\subseteq \V_{2.1}\subseteq
                 \V_{1.1}\subseteq \V_{0.1}\subseteq\V_{0.0}\,.
                                               \eqno (1.14)
                   $$
 One can consider also subspaces $\{\V_{s.0}\}$  of the
 space ${\V_{0.0}}$
                           $$
 \V_{4.0}\subseteq \V_{3.0}\subseteq \V_{2.0}\subseteq
                 \V_{1.0}\subseteq \V_{0.0},
            \quad \V_{s.1}\subseteq\V_{s.0}
                                               \eqno (1.15)
                   $$
 which correspond to  $\{\V_{s.1}\}$
 if we ignore the condition (1.12):
 The weakly $\G$-invariant Lagrangian $L$ belongs to
  $\V_{1.0}$ if $\d_i L=d\a_i+c_i$.
 It is easy to see that in this case
 $\d\a$ is also $2$-cocycle as in (1.4).
 $L\in\V_{2.0}$ if this cocycle is trivial,
  $L\in\V_{4.0}$ if $\a_i=\d_i f$ and
  $L\in\V_{3.0}$ if it differs from
  $\V_{4.0}$ on a closed form.  Lagrangians in
 $\V_{s.0}$ in general have time-dependent
 Noether currents (1.13).

  What can we say more about embeddings (1.14, 1.15)?
  Does weakly $\G$-invariant Lagrangian possesses generalized
 symmetries (1.2)?  Can it be reduced to $\G$-invariant?
 Does there exist Lagrangian which
 belongs to the space $\V_{s.0}$ and which does not
 belong to the space $\V_{s+1.0}$ or $\V_{s.1}$?
 If an answer is "no" what are the reasons of it.

To answer on these questions we establish the hierarchy in
the space of weakly  $\G$-invariant Lagrangians.
 This hierarchy relates the phenomena
discussed above with cohomologies groups
 of the Lie algebra $\G$ and the configuration space $M$.

 The scheme of our considerations is the following.
  We are fixing
 configuration space $M$ and finite-dimensional algebra $\cal G$
 of its transformations. Then we establish relations between
 weakly $\G$-invariant Lagrangians
 on $M$  and the
 cohomologies of algebra $\cal G$ and $M$.
 From considerations above we see that in the phenomena which
 we are investigating are interplaying two differentials $\d$
 and $\der$ where the differential $\d$
  corresponds to the symmetries
 and $\der$ is the prolongation of exterior differential
 which acts on Lagrangians.
 (It is variational derivative, whose action leads to Euler-Lagrange
 equation. See in details the Section II).
 These differentials as well as
differentials $d$ and $\d$ satisfy the conditions:
  $\d\sp 2=\der\sp 2=\der\d-\d\der=0$.
  We naturally come to the differential
  $Q=\der\pm\d$ which is strictly related with our problem.
 For example  to condition  $\d L=d\a$ in (1.2) corresponds
 the condition $Q(L,\a_i)=(\der L,0, w=\d\a)$.
To  redefinition of Lagrangian on a full derivative
 $L\rightarrow L+df$ corresponds the changing of
  the cochain $(L,\a_i)$ on coboundary:
     $(L,\a_i)\rightarrow(L,\a_i)+Qf=(L+df,\a_i+\d_i f)$.

  {\it It is the cohomology of the differential $Q$
 which allow us to reveal the
 relations between generalized symmetries
 of Lagrangians and cohomologies of the configuration space and
 the symmetries Lie algebra.}
 We do it in a following way.  Using
  a technique of spectral sequences
  we calculate the cohomology of $Q$ via cohomology of
 $\der$
 by modulo $\d$, then vice versa via cohomology of $\d$
 by modulo $\der$.
 Calculating cohomology of operator $Q$ in the first way
 we come to the spaces $K_s$ which are expressed in terms
 of cohomologies of Lie algebra and configuration space.
 On the other hand, calculating the same
 cohomology in the second way,
 we come  naturally to the space $\V_{0.0}$ of weakly
 $\G$-invariant Lagrangians and to its subspaces
   $\{\V_{s.\sigma}\}$ (1.14,1.15).
 Natural relations which arise between the results of
 calculations
 in the first and in the second way
 lead to the sequence of homomorphisms between
  the  spaces $\{\V_{s.\sigma}\}$ and $\{K_s\}$
 which define these spaces in a recurrent way
 via the kernels of corresponding homomorphisms.

 This construction establishes hierarchy in the space of
 weakly $\G$-invariant Lagrangians
 making links between the physical properties
 of Lagrangians
 and pure mathematical objects:
 The condition that Lagrangian belongs to some space
 $\V_{s.\sigma}$ and it {\it does not belong to
 the space}
 $\V_{s+1.\sigma}$ or $\V_{s.\sigma+1}$
 in terms of this hierarchy is reformulated
 to the condition that the value of the corresponding
 homomorphism on it, is not equal to zero.
 The problem of analyzing  the content of the spaces
  $\{\V_{s.\sigma}\}$ and their differences is reduced to the
 problem of calculating the corresponding homomorphisms.
 For example in the case if the space $K_s$ is trivial,
then $\V_{s-1.\sigma}=\V_{s.\sigma}$.
 In particular if all the spaces $K_s$ are trivial
 then all weakly
 invariant Lagrangians are invariant (up to a full derivative).

 The plan of the paper is the following.

 In the Section II
 we consider the complex of Lagrangians,
 clarify its relations with corresponding complex
 of differential forms.

 In the Section III we present the calculations
 of cohomology of the differential $Q$ of
 the double complex of cochains which are defined on the Lie
 algebra $\G$ and take values in the functions on $M$ and
 in Lagrangians of classical mechanics.
 Using the results of these calculations
 in the Section IV we establish hierarchy in the space
 of weakly invariant Lagrangians and consider some
 general properties of this
 hierarchy. It is the main result of the paper.
 In this Section from our point of view we consider
 also the hierarchy for Lagrangians polynomial in
 velocities.

 In the Section  V using this hierarchy
 we calculate the content of the subspaces $\V_{s.\sigma}$
  in (1.14, 1.15) for some special cases of configuration spaces
  and symmetries algebras. In particular
  we perform this analysis for $so(3)$, Poincar\'e and
  Galilean algebras.

 In the Section VI we give some motivations for the technique
  we used in this paper.

In Appendixes we give a brief sketch on
the notion of Lie algebra cohomology
and calculation of double complexes cohomology
via corresponding spectral sequences.

                          \bigskip
 \centerline {\bf II The complexes of
      Lagrangians and Differential Forms}
   \smallskip
  Let $M$ be an $n$-dimensional manifold (configuration space)
  and $\G$ be Lie algebra acting on it.
 It means that it is defined the homomorphism $\Phi$ from $\G$
 in the Lie algebra of vector fields
 on $M$:
                     $$
 \G\ni x{\buildrel \Phi\over\longrightarrow} {\t x}
   \,\,\hbox{(fundamental vector fields)}\,\colon\quad
               [{\t x},{\t y}]=\tilde {[x,y]}\,.
                                                    \eqno (2.1)
                      $$
   We denote this construction by $[\G,M]$ pair.

 Let $\Omega^q(M)$ be the space of differential $q$-forms on $M$.
The linear spaces $\Omega^q(M)$ for any given $q$
can be considered as modules on
 $\G$ if we define the action of algebra on forms via Lie derivatives
 along corresponding fundamental vector fields:
    $h\circ w={\cal L}_{\t h} w$.
 One can consider the $\G$-differential corresponding to
 this module structure
 and cohomologies spaces $H^q(\G,\Omega^q(M))$ which are
 $\G$-cohomologies with coefficients in $\Omega^q(M)$.
 (See Appendix 1).

 On the manifold $M$  endowed with the action of
 Lie algebra $\G$ one can also consider usual
 de Rham cohomologies $H^q(M)$ of the differential forms
complex  $\{\Omega^q,d\}$,
 where $d$ is exterior differential.
 One can naturally prolongate the
 action of exterior differential $d$ from the spaces $\Omega^q(M)$
($0$-cochains) on the spaces
 $C^p(\G,\Omega^q(M))=C^p(\G)\otimes\Omega^q(M)$ of
  $p$-cochains on the Lie algebra
  $\G$ with values in $\Omega^q$, taking values
 of $d$ on cochains in constants to be zero.
 The differentials $d$ and $\d$
 commute with each other:
      $d\d=\d d$
  and one can consider the corresponding double complex
   $\{C^p(\G,\Omega^q(M)),d,\d\}$.

 To include Lagrangians in a game we
  enlarge the complex $\{\Omega^q,d\}$
 of differential forms to the complex
 $\{\Lambda^q(M),\der\}$ of Lagrangians, following [\Vor].

 We define the space $\Lambda^q(M)$ of $q$-Lagrangians
 ($q\geq 1$) as the space of
 functions (Lagrangians) which depend on
  points $q^\mu$ of manifold $M$ and
 on derivatives

\noindent ${\p q^\mu\over\p\xi^\a},\dots,
 {\p^k q^\mu\over\p\xi^{\a^1},\dots,\p\xi^{\a^k}}$
 of an arbitrary but finite order $k$ of parameters
 $(\xi^1,\dots,\xi^q)$ which take values in
$q$-di\-men\-si\-onal
space ${\bf R}^q$.
 In the case $q=0$ we put $\Lambda^0(M)=\Omega^0(M)$
is the space of functions on $M$.
We say that Lagrangian has the rank $k$ if the highest degree
 of derivatives on whose it depends is
equal to $k$ and we denote by
 $\Lambda^q_k$ the subspace of $\Lambda^q$
 which contains $q$-Lagrangians of the rank $k$.
 The  Lagrangians of classical
 mechanics which we consider in the following Sections
 belong to $\Lambda^1_1$.

  If $L$ is the Lagrangian in $\Lambda^q(M)$ then to every
  map ($q$-dimensional path)

                   \line{\hss$
   q^\mu(\xi^1,\dots,\xi^q)\colon R^q\rightarrow M$\hss
                                                    (2.2)}
corresponds the integral
                       $$
           S_L([q(\xi)])=\int L
                \left(
    q^\mu(\xi),{\p q^\mu(\xi)\over\p\xi^\a},\dots,
  {\p q^\mu(\xi)\over\p\xi^{\a^1},\dots,\p\xi^{\a^k}}
                 \right)
               d\xi^1\dots d\xi^q \,.
                                           \eqno (2.3)
                       $$
   This defines the natural embedding
 of the space $\Omega^q(M)$   of differential $q$-forms
 in $\L^q_1(M)$:

\line{\hss$
 w=w_{i_1\dots i_q}(q)dq^{i_1}\wedge\dots\wedge dq^{i_q}
                     \longrightarrow
         L_w= n!w_{i_1\dots i_q}(q)
                {\p q^{i_1}\over\p\xi^1}\cdot\dots\cdot
              {\p q^{i_q}\over\p\xi^q}\,.
                       $\hss                         (2.4)}
  The integral $S_{L_w}([q(\xi)])$
 is equal to the integral of differential form $w$ over
 the surface which is the image
 of the map (2.2). It does not depend on
 choice of parametrization $q(\xi)$ of this surface.
 We say that
 Lagrangian $L_w$ corresponds to the differential form $w$
and later on we often will not differ $w$ and $L_w$.

 {\bf Remark } In general  for an arbitrary Lagrangian
 the l.h.s. of (2.3) is not correctly defined
on  images of maps (2.2). It can be considered as
 functional on embedded surfaces which does not depend on its
parametrization in a case if Lagrangian $L$
 is a {\it density}, i.e.
  under reparametrization
  $q(\xi)\rightarrow$  $q(\xi({\t\xi}))$,
   $L\rightarrow$  $L\cdot det(\p\xi/\p{\t\xi})$
 (see for example [\Kh,\Voron.]).
 The Lagrangians corresponding to differential
 forms are the special examples of densities.
    \smallskip

 To define the  complex
 of Lagrangians
 which is the generalization of
 de Rham complex  we consider following [\Vor]
 the differential
 $\der$,  using Euler-Lagrange equations of motion for
 the functional  (2.3):
                    $$
  \der\colon\, \L^q\rightarrow\L^{q+1},\quad
                        \der L
             \left(q,{\p q^\mu\over\p\xi^{\t a}},\dots,
 {\p q^\mu\over\p\xi^{{\t a}^1},\dots,\p\xi^{{\t\a}^k}}
             \right)=
         {\cal F}_\mu(L) {\p q^\mu\over \p\xi^{q+1}}\,.
                                            \eqno (2.5)
                      $$
 where
 $\t\a=(1,\dots,q,q+1),\a=(1,\dots,q)$ and
${\cal F}_\mu(L)$ are l.h.s. of
 Euler--Lagrange equations of the Lagrangian $L$, i.e.
  variational derivatives
 of the corresponding functional (2.3):
  $\,{\cal F}_\mu(L)= {\d\over\d q^a}S_L([q(\xi^\a)])$

  For example if $L\in\Lambda^q_1(M)$,
  $L=L(q,{\p q^\mu\over\p\xi^a})$ then
                     $$
      \der L\left
     (q,{\p q^\mu\over\p\xi^{\t a}},
 {\p^2 q^\mu\over\p\xi^{{\t a}}\p\xi^{{\t\beta}}}
             \right)=
                   \left(
                    {\p L\over\p q^\mu}-
            {\p^2 L\over\p q^\nu\p q^\mu_\a}
             {\p q^\nu\over\p\xi^\a}-
               {\p^2 L\over\p q^\nu_\beta\p q^\mu_\a}
              {\p^2 q^\nu\over\p\xi^\a\p\xi^\beta}
                     \right)
                        {\p q^\mu\over\p\xi^{m+1}}\,.
                                         \eqno (2.6)
                           $$

(In general $\der\L^q_k\subseteq \L^{q+1}_{2k}$)

 One can show  that as well as for exterior differential
$d$, $\der^2=0$ [\Vor] and consider cohomology
of the complex
                           $$
                      \{\L^q(M),\der\}\colon\quad
            {\Lambda^0(M)}{\buildrel \der\over\longrightarrow}
             {\Lambda^1(M)}{\buildrel \der\over\longrightarrow}
             {\Lambda^2(M)}{\buildrel \der\over
                 \longrightarrow}\dots\,.
                                                      \eqno (2.7)
                           $$
 From the definition of $\der$ and (2.4) it follows that:
 $L_{dw}=\der L_w$.
 The complex $\{\Omega^m(M),d\}$ of differential forms
 is subcomplex of the
 complex (2.7).

 The spaces  $\ \Lambda^q(M)$ of Lagrangians for any given $q$
 (and their subspaces $\Lambda^q_k(M)$ for any given $q$
 and $k$) as well as  $\Omega^q(M)$ can be
 naturally considered as modules
 on Lie algebra $\G$ if we define the
 action of Lie algebra
 elements on Lagrangians via Lie derivative:
if $x\in \G$ and $\t x=\Phi x=X^\mu(q)\p/\p q^\mu$ then
                  $$
              (x\circ L)={\cal L}_{\t x}L=
       X^\mu{\p L\over \p q^\mu}+
        D_\a X^\mu{\p L\over \p q^\mu_\a}+
         D_\beta D_\a X^\mu{\p L\over \p q^\mu_{\a\beta}}+\dots
                                                 \eqno (2.8)
                        $$
 where $D_\a={d\over d\xi^\a}=
 q^\mu_\a{\p\over q^\mu}+
 q^\mu_{\a\beta}{\p\over q^\mu_{\beta}}+\dots$
 is the full derivative.
 If a Lagrangian
 corresponds to differential form then (2.8) corresponds
 to usual Lie derivative: ${\cal L}L_w=L_{{\cal L}w}$.
 To the identity ${\cal L}_\eta w=dw{\cal c}\eta+d(w{\cal c}\eta)$
 for Lie derivative on forms
 corresponds the identity
    ${\cal L}_\eta L=\eta^\mu{\cal F}_\mu(L)+D_\a N^\a$
 which leads to Noether currents $N^\a$ in the case if
 ${\cal L}_\eta L=0$.

 Considering  $\G$-differential $\d$ corresponding to this module
 structure we come to the spaces $H^p(\G,\L^q_k(M))$ of
 $\G$-cohomologies with coefficients in $\L^q_k(M)$.

 In the same way like for differential
 forms one can prolongate the action of
 $\der$ on the  spaces $C^p(\G,\L^q)$ of $p$-cochains with
 values in $\L^q$ and consider the double complex
 $\{C^p(\G,\L^q),\der,\d\}$ because for Lagrangians
 $\der$ and $\d$ commute also.
   The complex
  $\{C^p(\G,\Omega^q),d,\d\}$ is embedded
 in this complex.

 The cohomology of the complex (2.7) evidently
 is different
 from  de Rham cohomology, but on the other
 hand

 {\bf Proposition 1}
 \footnote{$^*$}
  {The complex (2.7) differs from the standard variational complex
  (See for example [\Olv].) It was introduced by Th. Voronov
  in [\Vor] for the Lagrangians on superspace.
  This  complex
  and Proposition are
 useful in supermathematics where the concept of usual
 differential form is ill-defined [\Voron,\Khud].}

   1.If  Lagrangian $L$ is exact: $L=\der L^\prime$
 and it is a density (see the Remark above),
 then it corresponds to an exact differential form.

  2. If Lagrangian $L$ is closed and it
 depends only on first derivatives:
  $\der L=0, L\in\Lambda^q_1$, then it corresponds to
 closed differential form up to a constant
                           $$
                  L=L_w+c,\,dw=0\,.
                                                \eqno (2.9)
                           $$
   In the case if $L$ in (2.9) is a density then $c=0$.
  \smallskip
   The 2-nd statement immediately follows from
  (2.6) and the definition of the density. The
 1-st one we do not need here and we omit its proof.

  We use this Proposition to consider the following subcomplex
  $(\C^*,\der)$ of the complex (2.7), which will be of use in
this paper:
                        $$
  (\C^*,\der) \colon\qquad
            \Lambda^0(M){\buildrel \der\over\longrightarrow}
             \Lambda^1_1(M){\buildrel \der\over\longrightarrow}
                       \der\Lambda^1_1(M)
                         \longrightarrow 0
                                                    \eqno (2.10)
                             $$
  where as well as in (2.7)
 $\C^0=\L^0(M)$ is the space of functions on $M$;
   $\C^1=\L^1_1(M)$ is the
 space of Lagrangians
  $L(q^\mu,\dot q^\mu)$ of classical mechanics defined on
 the configuration space $M$,
  $\C^2$ is the subspace of coboundaries in
 $\L^2_2$.  It  contains elements
  corresponding to equations of motion of some
  Lagrangian from $\L^1_1$:
   $a\in \der\Lambda^1_1(M)$ iff there exists
  Lagrangian $L$ such that $a=\der L$.

 From the 2-nd statement of Proposition 1 it follows
 that cohomology of this truncated complex
 is strictly related with de Rham cohomology:
                     $$
             H^0(\C^*,\der)=H^0(M)\,,
             H^1(\C^*,\der)=H^1(M)+{\bf R},\,
                       H^2(\C^*,\der)=0\,.
                                                \eqno (2.11)
                   $$

  For our purposes it is useful to consider
 also the following modification of the complex
 (2.7). We consider the spaces $\{\overline {\L^q}\}$
 where
  $\overline {\L^q}=\L^q/{\bf R}$ if $q\geq 1$ and
  $\overline {\L^0}=\L^0=\Omega^0(M)$.  Elements of
 $\overline {\L^q}$ ($q\geq1$) are
  $q$-Lagrangians which are defined up to constants.
  We denote by $\overline L$ the equivalence class
 of Lagrangian $L$
  in $\overline \L$.
  One can consider
 instead the complex (2.7) the complex
                           $$
            \{\B{\L^q}(M),\B\der\}\colon\quad
     {\L^0(M)}{\buildrel \B\der\over\longrightarrow}
     {\B{\L^1}(M)}{\buildrel \B\der\over\longrightarrow}
     {\B{\L^2}(M)}{\buildrel \B\der\over\longrightarrow}\dots\,.
                                                   \eqno (2.12)
                           $$
  and correspondingly  the double complex
 $\{C^p(\G,\B{\L^q}),\B\der,\B\d\}$
  of $p$-cochains on $\G$ with values in $\B{\L^q}$.
 The differentials $\B\der$ and $\B\d$  are correctly
 defined
 in a natural way:
 $\B{\der} \bar\lambda\dot=\B{\der \lambda}$ and
 $\bar\d\bar\lambda=\B{\d\lambda}$ where $\B\lambda$
 is equivalence class
 in $C^*(\G,\B{\L^*})$ of the cochain
 $\lambda$ in $C^*(\G,\L^*)$.
 The differential $\B\der$
 does not differ essentially
 from $\der$: If $\lambda$ is
 cochain with values in Lagrangians then
 it is easy to see that
                           $$
     \B\der\B\lambda=0,\iff \der\lambda=0\,
                                                 \eqno (2.13)
                  $$

  To  (2.10) corresponds the subcomplex
                        $$
  (\B{\C^*},\der) \colon\qquad
           \Lambda^0(M)
      {\buildrel \B\der\over\longrightarrow}
      \overline{\Lambda^1_1}(M)
   {\buildrel \B\der\over\longrightarrow}
          (\der\overline{\Lambda^1_1}(M))
                \longrightarrow 0
                                                 \eqno (2.14)
                             $$
 of the complex (2.12).
 From  (2.13) it follows that
  for the truncated complex $\B{\C^*}$
                     $$
              H^0(\B{\C^*},\der)=H^0(M),\,
              H^1(\B{\C^*},\der)=H^1(M),\,
           H^2(\B{\C^*},\der)=0\,.
                                                   \eqno (2.15)
                   $$
We avoid here an appearance of non-pleasant constants
like in (2.11).
The difference between the complex
 $\{C^p(\G,\B{\L^q}),\B\der,\B\d\}$   and
the complex $\{C^p(\G,\L^q),\der,\d\}$
 becomes
 non-trivial at least on the level of $1$-cochains.
 It corresponds to the difference of time independent and
 time dependent Noether charges.
  (See for e.g. the Example 1 in the Section V.)

  Finally we want to note that to every Lagrangian  $L$
 on $M$ corresponds the density $A_L$ on the space
  $\hat M=M\times \{\hbox{space of parameters}\}$. (It is so called
formalism where fields and space variables are on an equal footing
 [\Sch]).  To the functional (2.3) corresponds the
integral of the density over the surface in
 $\hat M$ which is the graph
 of the map (2.2).
 For example to Lagrangian $L(q^\mu,{dq^\mu\over dt})$
 of classical mechanics one can correspond the density
                          $$
           A_L\left(
         q^\mu,{dq^\mu\over d\tau},{dt\over d\tau}
                  \right) =
           L\left(q^\mu,{dq^\mu\over d\tau}
            \big/{dt\over d\tau}\right)
          \cdot {dt\over d\tau};\quad{\rm if}\,
      \tau\rightarrow \tau^\prime(\tau),\,{\rm then}\,
            A_L\rightarrow {d\tau\over d\tau^\prime}A_L\,.
                                                  \eqno (2.16)
                             $$
 To a path $q^\mu(t)$ corresponds curve $(q^\mu(\tau),t(\tau))$
 and   $S_L([q(t)])=S_{A_L}([q(\tau),t(\tau)])$ for
 any parametrization $q(\tau)$.
  It is easy to see that for densities $A_L$
 the difference between complexes (2.10) and (2.14)
 is  removed.
 To redefinition of Lagrangian $L$ on the constant $c$ corresponds
 redefinition of $A_L$ on the form $cdt$.)

           \bigskip
      \centerline {\bf III Cohomology of Lagrangians
        Double Complex and}
      \centerline {\bf and its Spectral Sequences.}
 \smallskip
 Now using the technique briefly described in the
previous Section and in Appendix 2
we investigate systematically the problem
which we considered in Introduction.

 We study simultaneously two double complexes,
 the double complex  $(E^{*.*},\der,\d)$  of
   cochains on $\G$ with values in
   the spaces of the complex $\C^*$ defined by (2.10),
    $\{E^{p.q},\der,\d\}$

  \noindent $=\{C^p(\G,\C^q),\d,\der)\}$
 and the double complex  $(\B{E^{*.*}},\B\der,\B\d)$  of
   cochains on $\G$ with values in
   the spaces of the complex $\B\C^*$ defined by (2.14),
    $\{\B {E^{p.q}},\B\der,\B\d\}=
  \{C^p(\G,\B{\C^q}),\B\der,\B\d)\}$.

 The complex  $(E^{*.*},\der,\d)$ is
 represented by the following
 table
                             $$
                            \matrix
                                {
         \L^0(M) &\drightarrow &\L^1_1(M) &\drightarrow
               &\der\L^1_1(M) &\drightarrow &0 \cr
          \d \downarrow &   &\d\downarrow       &   &\d\downarrow
                                   &   &         \cr
       C^1(\G, \L^0(M)) &\drightarrow &C^1(\G,\L^1_1(M)) &\drightarrow
                &C^1(\G, \der\L^1_1(M)) &\drightarrow &0\cr
           \d\downarrow &   &\d\downarrow       &   &\d\downarrow
                          &   &\cr
       C^2(\G, \L^0(M)) &\drightarrow &C^2(\G,\L^1_1(M)) &\drightarrow
                &C^2(\G,\der\L^1_1(M)) &\drightarrow &0\cr
           \d\downarrow &   &\d\downarrow       &   &\d\downarrow
                    &   &\cr
           \cdot &   &\cdot       &   &\cdot           &   &\cr
           \cdot &   &\cdot       &   &\cdot           &   &\cr
           \cdot &   &\cdot       &   &\cdot           &   &\cr
                        }
                                                     \eqno (3.1)
                          $$
 (The table represented the complex $(\B{E^{*.*}},\der,\d)$
differs from (3.1) by putting the
        "bar"s in corresponding places.

The differential $Q$ of the complex (3.1) is equal to
                          $$
                       Q=(-1)^q\d+\der,\,
                          \hbox {for the complex $E^{*.*}$}
                                                      \eqno (3.2)
                          $$
 and correspondingly  $\B Q=(-1)^q\B\d+\B\der$
 for the complex $\B{E^{*.*}}$.

  The problem of weakly invariant Lagrangians classification
 can be reformulated in terms of these double complexes.

  For this purpose we consider their spectral sequences
      $\{E^{*.*}_r\},\{\B{E^{*.*}_r}\}$ and
 transposed spectral sequences
      $\{\,^tE^{*.*}_r\},\{\B{\,^tE^{*.*}_r}\}$.

   The relations between
   $\{\,^tE^{*.*}_r\}$ and $\{E^{*.*}_r\}$  lead to the hierarchy
 in the space of weakly $\G$-invariant Lagrangians
 with time-independent Noether charges,
   the relations between  $\{\B{\,^tE^{*.*}_r}\}$
and $\{\B{E^{*.*}_r}\}$  lead to the hierarchy
in the space of weakly $\G$-invariant Lagrangians
 with time-dependent Noether charges
  and  the relations between  $\{E^{*.*}_r\}$
and $\{\B{E^{*.*}_r}\}$  lead to the relations
between these two hierarchies.

  We denote by $\V_{0.0}$ (see Introduction) the subspace
 of weakly $\G$-invariant Lagrangians in the space $E^{0.1}$,
i.e. Lagrangians whose motions equations l.h.s. are $\G$-invariant:
                  $$
     \V_{0.0}=\{L\colon\, L\in \L_1^1\,{\rm and}\,
               \d\der L=0\}\,.
                                       \eqno (3.3)
                 $$
 One can see that the cochain
${\bf f}=(\der L,0,0)$
 is the cocycle of differential $Q$ iff $L\in \V_{0.0}$.
 The cohomology class $[(\der L,0,0)]$ of this cocycle belongs
 to $H^2(Q)$.
 If we express the cohomology of
 differential $Q$  via the stable terms of transposed spectral
 sequence $\{\,^tE^{*.*}_r\}$, i.e. calculating $H^*(Q)$
 in perturbation theory, considering in (3.1)
 the differential $\d$ as zeroth order approximation
 for the differential $Q$, we see that
$[\der \V_{0.0}]_\infty=\,^tE^{0.2}_\infty$ is the subspace of
$H^2(Q)$. On the other hand
if we express the cohomology of
differential $Q$  via the stable terms of spectral
 sequence $\{E^{*.*}_r\}$, i.e. calculating $H^*(Q)$
 in perturbation theory, considering in (3.1)
 the differential $\der$
 as zeroth order approximation, we express $H^2(Q)$ in terms
 of $\{E^{p.2-p}_\infty\}$.
 The relations between the space $^t E^{0.2}_\infty$ and
 the spaces $\{E^{p.2-p}_\infty\}$
 lead to the relations between the space of weakly
 $\G$-invariant Lagrangians and cohomologies groups of $\G$
 and $M$.

\medskip
           \centerline{{\it The spaces } $\{E^{p.q}_r\}$ and
             $\{\B{E^{p.q}_r}\}$}
\smallskip
 We pay more attention on the calculations for the spaces
 $\{E^{*.*}_r\}$.
 The calculations for the spaces $\{\B{E^{*.*}_r}\}$
can be performed analogously using (2.13).

\noindent The spaces  $\{E^{p.q}_1\}$ are equal to the cohomologies of
operator $\der$:
 $E^{p.q}_1=H(\der,E^{p.q})$.

\noindent(See Appendix 2).

 From (2.11) and (2.15) it immediately follows that
                             $$
                            \matrix
                                {
                 &E^{*.*}_1 &&      &\B{E^{*.*}_1} &\cr
                 &&&&&\cr
   {\bf R}       &H^1(M)\oplus{\bf R}   &0&\phantom{qwert}
   {\bf R}       &H^1(M)                &0\cr
   C^1(\G)       &C^1(\G,H^1(M)\oplus{\bf R})
    &0& \phantom{qwert}
   C^1(\G)         &C^1(\G,H^1(M))       &0\cr
   C^2(\G)        &\cdots   &0&\phantom{qwert}
   C^2(\G)                  &\cdots                &0\cr
   \cdots         &\cdots      &0&\phantom{qwert}
   \cdots         &\cdots          &0\cr
                           }
                                                   \eqno (3.4)
                          $$
 Hereafter we identify the differential forms with Lagrangians
corresponding to them by (2.4) and
 the differential $\der$ on these Lagrangians with differential
$d$ on forms.

  In the columns of $E^{*.*}_1$ acts operator $d_1$
  which is generated by
 $\d$ and according to (A2.13)
$E^{p.q}_2=H(E^{p.q}_1,d_1)$.
It is easy to see that  $E^{p.0}_2=H^p(\G)$
is $p$-th cohomology group
of the Lie algebra $\G$ with coefficients in ${\bf R}$.

 Now we prove that $E^{0.1}_1=E^{0.1}_2$. Indeed if
 $c\in E^{0.1}_1$ is a constant ($c\in {\bf R}$)  then $d_1 c$
is evidently equal to zero.
To prove that $d_1 H^1(M)=0$ we consider the following
homomorphism $\pi$ from the space of differential $1$-forms into the
space  of $1$-cochains on $\G$ with values in
 functions on $M$ (the space $\L^0(M)$):

 \line{\hss $(\pi w)(h)= w{\cal c}\t h$,\hss          (3.5)}
 where $\t h$ is the fundamental vector field $\Phi h$ corresponding
 to the element $h$ of the Lie algebra $\G$ by (2.1).

From the standard formulae of differential geometry it follows that

  \line{\hss if $dw=0$ then $\d\pi w=0$ and
     $\d w=d\pi w$.\hss               (3.6)}
Hence for the cohomology class
 $[w]$ in $H^1(M)$ $d_1[w]=[\d w]=[d\pi w]=0$ in $E^{1.1}_1$.
 Hence
 $Z^{0.1}_1=E^{0.1}_1$ and $E^{0.1}_1=E^{0.1}_2$  because
 $B^{0.1}_1=0$.

   Now we calculate
  $E^{1.1}_2$. If $[c]_1\in E^{1.1}_1$ then
                       $$
        c=\sum_\lambda t^{(\lambda)}\otimes w^{(\lambda)}
          +t^\prime +d\a
                                                  \eqno (3.7)
                       $$
 where $t,t^\prime$ belong to $C^1(\G)$ (are constants),
 the set $\{w^{(\lambda)}\}$ of differential closed
 $1$- forms constitutes a basis
 in the space $H^1(M)$ of
 $1$-cohomology and $\a$ is some element from $E^{1.0}$.
   The straightforward calculations using (3.5, 3.6) give that
                            $$
                      d_1 [c]_1=
        \sum_\lambda[ \d t^{(\lambda)}\otimes w^{(\lambda)}
          +\d t^\prime +d(\dots)]=0
      \Rightarrow \d t^{(\lambda)}=0\,\,
               {\rm and}\,\, \d t^\prime =0\,.
                                                 \eqno (3.8)
                          $$
  On the other hand
 coboundaries
 in  $E^{1.1}_1$ are equal to zero
 because $E^{0.1}_1=E^{0.1}_2$. Hence
 from   eq. (3.7) it follows that
$E^{1.1}_2=H^1(M)\otimes H^1(\G)\oplus H^1(\G)$.
 (In the case of complex $\B{E^{1.1}_1}$,
 $t^\prime$ in (3.7)
is equal to zero and from (2.13)  it follows that
  (3.8) holds also.)

  We arrive at the following tables
                             $$
                            \matrix
                                {
                 &E^{*.*}_2 &&      &\B{E^{*.*}_2} &\cr
                 &&&&&\cr
   {\bf R}       &H^1(M)\oplus{\bf R}   &0&\phantom{qwert}
   {\bf R}       &H^1(M)                &0\cr
   H^1(\G)       &H^1(\G)\otimes H^1(M)\oplus{\bf R}
    &0& \phantom{qwert}
   H^1(\G)         &H^1(\G)\otimes H^1(M)       &0\cr
   H^2(\G)        &\cdots   &0&\phantom{qwert}
   H^2(\G)                  &\cdots                &0\cr
   H^3(\G)         &\cdots      &0&\phantom{qwert}
   H^3(\G)         &\cdots          &0\cr
                           }
                                                   \eqno (3.9)
                          $$

  \smallskip

  One can show that the spaces
 $\{E_2^{p.q}\}$ in (3.9)
 which are of interest for us ($p+q\leq 2$) are stable:
 $E_2^{p.q}=E_3^{p.q}=\dots=E_\infty^{p.q}$.
 (The same for $\{\B{E_2^{p.q}}\}$.)

  It is evident without any calculations for the spaces
 $E^{0.0}_2$, $E^{1.0}_2$
 because differentials $d_2$ which acts on these spaces
 goes out of the table and
 the boundaries are zero by the same reasons.
  The spaces $E_2^{0.1}$ and $E_2^{2.0}$
 are stable
 because
 the differential $d_2$ acting from the space
 $E_2^{0.1}$ into the  space $E_2^{2.0}$ transforms it to zero.
 It follows from eq. (3.5):
$d_2 [w]=[Q (w, \pi w)]=[\d\pi w]=0$.
The same arguments lead to the stability of the space
$E^{1.1}_2$.
 One can perform the analogous considerations for the spaces
 $\{\B{E_2^{p.q}}\}$.

   Hence the tables (3.9) establish the relations
 between the spaces $H^m(Q)$, $H^m(\B Q)$ ($m=0,1,2$) and
 the spaces $E_\infty^{p.q}$,
 $\B{E_\infty^{p.m-p}}$ correspondingly, according to  eq. (A2.11).

  $H^0(Q)=H^0(\B Q)={\bf R}$.
  Considering the first "antidiagonal"
 $\{E^{0.1}_\infty,E^{1.0}_\infty,\}$  in (3.9)  we see
from (A2.11) that

  \line{$\hss H^1(\G)\subseteq H^1(Q)\quad {\rm and}
                 \quad H^1(M)\oplus{\bf R}=
                  H^1(Q)\big/H^1(\G)$.\hss
                                            (3.10)}
 These relations define canonical projection $p_1$ of
 $H^1(Q)$ on  $H^1(M)\oplus {\bf R}$
 and isomorphism $\iota_1$ of ${\bf ker}p_1$ on
 $H^1(\G)$: If ${\bf L}=(L,\a)$ is a cocycle of $Q$
 then $L=w+c$ where $w$ is a closed
 form and $c$ is a constant and  $p_1([{\bf L}])=[w]+c$.
 If $c=0$ and $w=df$ then $\a-\d f$ is 1-cocycle in constants
 which is equal to  $\iota_1([{\bf L}])$.

   Using the homomorphism (3.5) one can
  establish also the isomorphism

\line{\hss$
   H^1(M)\oplus H^1(\G)\oplus {\bf R}\longrightarrow
                H^1(Q)\colon\quad
           [w]+t+c\longrightarrow [w+c,t+\pi w]$\hss
                                               (3.11)}
 which corresponds to (3.10) and
 splits $H^1(Q)$ on components.

  The analogous considerations  for the table $\B{E^{*.*}_2}$
lead  to formulae analogous to (3.10, 3.11):
       $H^1(\G)\subseteq H^1(\B Q)$ and
  $H^1(M)=H^1(\B Q)\big/H^1(\G)$;
 $H^1(M)\oplus H^1(\G)=H^1(\B Q)$.

Considering in the same way the second "antidiagonal"
 $\{E^{0.2}_\infty,E^{1.1}_\infty,E^{2.0}_\infty\}$
 in (3.9)  we see that
                   $$
 H^2(\G)\subseteq H^2(Q)\quad{\rm and}\quad
 H^1(M)\otimes H^1(\G)\oplus H^1(\G)=H^2(Q)\big/H^2(\G)\,.
                                           \eqno (3.12)
                    $$
  These relations define canonical projection

    \line{\hss$p_2\colon\quad
 H^2(Q)\longrightarrow H^1(M)\otimes H^1(\G)+H^1(\G)$
                                          \hss      (3.13)}
 and on the kernel of $p_2$ the isomorphism

    \line{\hss$
\iota_2\colon\quad {\bf ker} p_2\longrightarrow H^2(\G)$.\hss
                                                 (3.14)}
   We consider now (3.13) and (3.14) in components.

  Let  ${\bf f}=[{\cal F},\lambda,f]\in H^2(Q)$
 be a  cohomology class of
 cocycle $({\cal F},\lambda,f)$: $Q({\cal F},\lambda,f)=0$.
  $\der\lambda=-\d{\cal F},\d\lambda=df,\d f=0$.
    (${\cal F}\in E^{0.2},\lambda\in E^{1.1},f\in E^{0.2}$).
 The space $E^{0.2}$ contains only coboundaries, so
 cocycle  $({\cal F},\lambda,f)$ is cohomological to
 $(0,\lambda^\prime,f)$ where $\lambda^\prime=\lambda-\d L$
($L\colon\,{\cal F}=\der L$). $\der\lambda^\prime=0$, so
from  Proposition 1 it follows that
 $1$-cochain $\lambda^\prime$ takes values
 in closed differential $1$-forms $+$ constants:

   \line{\hss $\forall h\in\G\quad \lambda^\prime(h)=w(h)+t(h)$.
                                               \hss (3.15)}
 Using (3.7,3.8) we see that to $\lambda^\prime$
 corresponds element of $H^1(M)\otimes H^1(\G)\oplus H^1(\G)$ which
 is nothing but $p_2({\bf f})$.

  In the case if
   $p_2({\bf f})=0$ then it means that
 $\lambda^\prime=d\a$ where $\a\in  E^{1.0}$
 and the cocycle  $(0,\lambda^\prime,f)$ is cohomological to
 a cocycle  $(0,0,f-\d\a)$.  $d(f-\d\a)=0$ so
   $f-\d\a$ is cocycle in $Z^2(\G)$. The cohomology
 class of $f-\d\a$ in $H^2(\G)$ is nothing but $\iota_2({\bf f})$.

 The analogous considerations for the second "antidiagonal"
 in the table $\B{E^{*.*}_2}$ lead to the analogous conclusions
 for $H^2(\B Q)$:
   $H^2(\G)\subseteq H^2(\B Q)$ and
 $H^1(M)\otimes H^1(\G)=$
 \break$H^2(\B Q)\big/H^2(\G)$;
  $\B{p_2}\colon\,H^2(\B Q)
 \longrightarrow H^1(M)\otimes H^1(\G)$.
 On the kernel of $\B{p_2}$ is defined
 isomorphism $\B{\iota_2}\colon\,\quad {\bf ker} \B{p_2}
      \longrightarrow H^2(\G)$.

  From the considerations above we see
 that natural relations between complexes

 \noindent
 $(E^{*.*},\der,\d),(\B{E^{*.*}},\B\der,\B\d)$
 lead to isomorphisms
                          $$
           H^1(Q)=H^1(\B Q)\oplus {\bf R},\quad
           H^2(Q)=H^2(\B Q)\oplus H^1(\G)\,.
                                 \eqno (3.16)
                         $$
The decomposition of $H^2(Q)$
 defines  the projection

   \line{\hss $\sigma\colon\, H^2(Q)\rightarrow H^1(\G)$.
                                               \hss (3.17)}
 $\sigma({\bf f})$ is equal to the element of
 $H^1(\G)$ in the r.h.s. of the eq. (3.15).
 This projection will be useful for extracting
 Lagrangians whose Noether charges are time independent
 in the space $\V_{0.0}$ of weakly invariant Lagrangians.

\smallskip
 Now we return again to the complex
  (3.1) and express the cohomologies
of $H(Q)$ and $H(\B Q)$ in terms of transposed spectral sequences
        $\{\,^t E^{*.*}_r\}$ and
        $\{\B{\,^t E^{*.*}_r}\}$.

                              \smallskip

 For constructing $^t E^{*.*}_1$ and
 $\{\B{\,^t E^{*.*}_r}\}$
 we have to consider
 as zeroth order approximation
 the cohomology of vertical differential $\d$:
    $\{\,^t E^{*.*}_1\}=H(E^{*.*},\d)$ and
    $\{\B{\,^t E^{*.*}_1}\}=H(\B{E^{*.*}},\d)$.
    We arrive at the tables
                          $$
                            \matrix
                                {
             &\,^tE^{*.*}_1 &&    &\B{\,^tE^{*.*}_1} &\cr
                 &&&&&\cr
   \L^0_{inv}   &{\L^1_1}_{inv}       &\der \V_{0.0}&\phantom{qwert}
   \L^0_{inv}   &\B{\L^1_1}_{inv}    &\B\der \B \V_{0.0}\cr
   \Hs       &H^1(\G,\L^1_1)
    &\cdots& \phantom{qwert}
   \Hs        &H^1(\G,\B{\L^1_1})       &\cdots\cr
   H^2(\G,\L^0(M))           &\cdots            &\cdots&\phantom{qwert}
   H^2(\G,\L^0(M))           &\cdots            &\cdots\cr
   \cdots         &\cdots      &\cdots&\phantom{qwert}
   \cdots         &\cdots          &\cdots\cr
                           }
                                                   \eqno (3.18)
                          $$
 Here $\L^0_{inv}=C^0(\G,\L^0(M))$ is the space of
the functions on $M$
which are invariant under the action of the Lie algebra $\G$.
The  ${\L^1_1}_{inv}$ is the space of $\G$-invariant Lagrangians from
 $\L^1_1$. The space $\B{\L^1_1}_{inv}$  in the right table
 contains the classes (Lagrangians factorised by constants)
 whose variation under $\G$ symmetry transformations produces
  $\G$-cochain with values in constants:
 $\B\L\in \B{\L^1_1}_{inv}\Leftrightarrow
 \B\d\B\L=0\Leftrightarrow \d_i\L=t_i$. These Lagrangians
 have linear time dependent Noether charges
 (see (1.13)).  The space $\der \V_{0.0}$  is the image under
 differential $\der$ of the subspace $\V_{0.0}$ of weakly
 $\G$-invariant Lagrangians (see (3.3)).
 From  (2.13) it follows that
   $\B{\,^tE^{0.2}}=\B{\der \V_{0.0}}$ also.

  The differential $\,^td_1$ which is
  generated by $\der$ acts in rows of
  the table $\,^tE^{*.*}_1$ (compare
 with the table (3.4)).
For $\,^tE^{*.*}_2=H(\,^tE^{*.*}_1,\,^td_1)$ we obtain
                          $$
          \,^tE^{*.*}_2=
                            \matrix
                                {
   {\bf R}   &H^1_{inv}(M)      \oplus{\bf R}
            &\der \V_{0.0}\big/(\der {\L^1_1}_{inv})\cr
        \,^tE^{1.0}_2
         &\cdots       &\cdots\cr
       \cdots           &\cdots            &\cdots\cr
                           }
                                                   \eqno (3.19)
                          $$
 $H^1_{inv}(M)$ is the space of closed
 $\G$-invariant differential
 $1$-forms factorised by the differentials of
 $\G$-invariant functions.

The analogous table one can consider for
 $\B{\,^tE^{*.*}_2}$.

 The space $\,^tE^{1.0}_2$ in (3.19) is the subspace of
 $\Hs$. It contains the classes $[\a]$ from
 $\Hs$  for whose the eq. $d\a=\d L$ has the solution.
(Compare with (1.2)). We see that
 the table (3.19) is not stable
in the spaces which we
are interesting in
 because the differential $\,^td_2$ acting
 from $\,^tE^{1.0}_2$ in $\,^tE^{0.2}_2$ is not trivial:
 $\,^td_2[\a]=\,^t[\der L]_2$.
 The next table
 $\,^tE^{*.*}_3=H(\,^tE^{*.*}_2,\,^td_2)$ is stable in
the spaces  which we are interesting in:
                          $$
                 \,^tE^{*.*}_3=
                            \matrix
                                {
   {\bf R}   &H^1_{inv}(M)\oplus{\bf R}
            &{\der \V_{0.0}\big/(\der {\L^1_1}_{inv})\over
            {\bf Im}(^td_2\,^tE^{1.0}_2)}\cr
               \,^tE^{0.1}_3
         &\cdots       &\cdots\cr
       \cdots           &\cdots            &\cdots\cr
                           }
                                                   \eqno (3.20)
                          $$
From the general properties of spectral sequences
 it follows that in (3.20) $\,^tE^{0.2}_3=\,^tE^{0.2}_\infty$
{\it is the subspace in} $H^2(Q)$ and the space
$\,^tE^{1.0}_3=\,^tE^{1.0}_\infty$
 (which is the subspace of $\,^tE^{1.0}_2$)
 is the factorspace  of $H^1(Q)$ by
  the space $\,^tE^{0.1}_3=H_{inv}(M)\oplus {\bf R}$
  (compare with (3.10)).
 Hence from  the decomposition (3.11) of $H^1(Q)$
it follows that
                     $$
      \,^tE^{1.0}_3=(H^1(M)\oplus H^1(\G))\big/H^1_{inv}(M)\,.
                                         \eqno (3.21)
                      $$
  In (3.21) $H^1_{inv}(M)$
 is considered as naturally embedded in
     $H^1(M)\oplus H^1(\G)$. If $w\in H^1_{inv}(M)$ is trivial
  in $H^1(M)$  then $\d f\in H^1(\G)$  where $w=df$.

Performing the corresponding calculations
 for the table  $\B{\,^tE^{*.*}_3}$ one has to
 put the "bar"s in $\,^tE^{0.2}_3$,
  the space
$\,^tE^{0.1}_3$  has to be changed on
  $\B{\,^tE^{0.1}_3}=H^1(M)_{inv}$. The spaces
  $\,^tE^{1.0}_1$ and $\B{\,^tE^{1.0}_1}$
as well as the spaces
$\,^tE^{1.0}_3$ and $\B{\,^tE^{1.0}_3}$ coincide but
on the other hand
 $\,^tE^{1.0}_2\subseteq\B{\,^tE^{1.0}_2}$.

  In the tables (3.18--3.20)
  every space $\,^tE^{1.0}_r$ is the
 subspace of previous one and correspondingly every space
  $\,^tE^{0.2}_r$ is the factorspace of previous one.
 We denote by $\Pi_r$ the homomorphism
which put in correspondence to every
weakly $\G$-invariant Lagrangian its equivalence class
 in the space $\,^tE^{0.2}_r$:
                          $$
   \Pi_r\colon\quad \V_{0.0}\rightarrow \,^tE^{0.2}_r,\quad
     \Pi_r(L) =\,^t[\der L]_r.\quad \forall L\in \V_{0.0}\quad
   {\bf Im}\Pi_3\subseteq H^2(Q).
                                    \eqno(3.22)
                           $$

   Analogously
   $\B\Pi_r\colon\quad \B \V_{0.0}\rightarrow \B{\,^tE^{0.2}_r}$.

 Comparing the content of
the spaces $\{\,^tE^{1.0}_r\}$  and $\{\,^tE^{0.2}_r\}$
in the transposed spectral sequences  (3.18--3.20)
 with the results above for spectral
sequence $\{E^{*.*}_r\}$ we come to
  \smallskip
   {\bf Proposition 2}

  a) To weakly $\G$-invariant Lagrangians
correspond elements in the space $\,^tE^{0.2}_3$,
 i.e.
 in $H^2(Q)$. Thus to these Lagrangians via
 homomorphisms $p_2$ and $\iota_2$ (3.13,3.14)
 correspond elements in $E^{1.1}_2$ or in $E^{2.0}_2$.

  b) To weakly $\G$-invariant Lagrangians
whose image in the space
$\,^tE^{0.2}_3$ is equal to zero: $\Pi_3(L)=0$,
 correspond elements in $\,^tE^{0.2}_2$
 which belong to the image of the differential
 $\,^td_2$. Thus to these Lagrangians
 correspond elements in $\,^tE^{1.0}_2$
which are defined up to
the space $\,^tE^{1.0}_3$
defined by (3.21), which is the
kernel of this differential.

 c) The space $\,^tE^{1.0}_3$  is related with
    weakly $\G$-invariant Lagrangians
whose image in the space
 $\,^tE^{0.2}_2$ is equal zero: $\Pi_2(L)=0$.

 The analogous statement is valid for the spaces
    $\{\B{\,^tE^{*.*}_r}\}$.
  \smallskip
 In the next section using this Proposition we
 establish the hierarchy in the space of
  weakly $\G$-invariant Lagrangians.

    \bigskip
   \centerline  {\bf IV The calculation of the hierarchy}
         \medskip

        Now using the calculations of the previous section
  for a given  pair $[\G,M]$ we
    establish the hierarchy in the space
  of weakly $\G$-invariant Lagrangians.

 Let  $\U$ be an arbitrary subspace in the space
 $\L^1_1(M)$ of the classical mechanics Lagrangians
on $M$.

 Let $\U_{0.0}$ be the subspace
 of weakly $\G$-invariant Lagrangians in $\U$:
 $\U_{0.0}=\V_{0.0}\cap \U$, where $\V_{0.0}$
 is the subspace
 (3.3) of the all weakly $\G$-invariant Lagrangians in
$\L^1_1(M)$.
 From the Proposition 1 and
  (3.3) it follows that  for an arbitrary  $L$ in $\U$
 the condition that $L\in \U_{0.0}$ is equivalent to the condition
 that the cochain $\d L$ takes values in
closed differential forms $+$
 constants:
                           $$
                       \matrix{
      \d\der L=0\Leftrightarrow
      \d_i L=w_{i\mu}\dot q^\mu+t_i\,\,
         {\rm and}\,\,dw_i=dt_i=0\,.\cr
           \hbox {(Compare with (1.11).)}\cr}
                                           \eqno (4.1)
                             $$
   $\d_i L$ is the value of the cochain $\d L$ on the basis vector
 $e_i$ of the Lie algebra $\G$.
  (As always we identify differential forms with
   Lagrangians corresponding to them by (2.4))

  Using the homomorphism $\Pi_3$ defined by (3.22)
 and the projection homomorphism (3.17)
 $\sigma_2$ of $H^2(Q)$ on $H^1(\G)$
 we consider the following composed homomorphism:
 $\Psi=\sigma\circ\Pi_3\colon\,\U_{0.0}
\rightarrow H^2(Q)\rightarrow K_0=H^1(\G)$.
 In the components according to (3.15),
$\Psi_i(L)=t_i$
where $t_i$ is defined by (4.1).
 We denote by $\U_{0.1}$ the kernel of this homomorphism.
 In the case $\U=\L^1_1(M)$ it is just the space $\V_{0.1}$
in (1.14) defined by the condition (1.12).

 Now on the subspaces of $\U_{0.1}$ and on the subspaces of
   $\U_{0.0}$ using the Proposition 2 we define in the
  recurrent  way the homomorphisms
   $\{\phi_s\}$ and correspondingly $\{\B\phi_s\}$  such that
 every homomorphism is defined on the kernel of previous one.
   Moreover the definition spaces for these homomorphisms
  will be related via the homomorphism
  $\Psi$.

  Using the statement a) of the Proposition 2
we consider the  composed  homomorphisms
  $\phi_1=p_2\circ\Pi_3\colon \U_{0.0}\rightarrow H^2(Q)
\rightarrow H^1(M)\otimes H^1(\G)\oplus H^1(\G)$ and
  $\B\phi_1=\B p_2\circ\B\Pi_3\colon \U_{0.0}\rightarrow  H^2(\B Q)
\rightarrow K_1=H^1(M)\otimes H^1(\G)$.
 From (3.15, 16) it follows that
 the restriction of $\B\phi_1$ on the  subspace $\U_{0.1}$ coincides
 with $\phi_1$.
 We denote by $\U_{1.0}$ the kernel of the homomorphism
 $\B\phi_1$  and by $\U_{1.1}$ the kernel of the homomorphism $\phi_1$.
 $\U_{1.1}$  is also the kernel of
 homomorphism $\Psi$ restricted on $\U_{1.0}$ .
 On the spaces  $\U_{1.1}$ and $\U_{1.0}$ one can consider composed
  homomorphisms
  $\phi_2=\iota_2\circ\Pi_3\colon \U_{1.1}\rightarrow H^2(Q)
\rightarrow K_2=H^2(\G)$ and
  $\B\phi_2=\B\iota_2\circ\B\Pi_3\colon \U_{1.0}\rightarrow H^2(\B Q)
 \rightarrow K_2$ correspondingly. $\B\phi_2$ evidently
  coincides with $\phi_2$ on $\U_{1.1}$.

 For example if for Lagrangian $L$ in $\U$ the condition (4.1) is
satisfied, i.e. $L\in \U_{0.0}$, then
 $\B\phi_1(L)$ is equal to the cohomology class of
 $w_{i\mu}dq^\mu$ in $H^1(\G)\otimes H^1(M)$ defined by
 (3.7);
 $L\in\U_{1.0}$  iff $\{w_{i\mu}dq^\mu\}$  are exact forms.
  In this case $\B\phi_2(L)$ is equal to the
  cohomology class in $H^2(\G)$
 of the cocycle $f_{ij}=(\d\a)_{ij}$ where $d\a_i=w_i$.
  If $t_i=0$ also then $L\in\U_{1.1}$.

 We denote by $\U_{2.0}$ the kernel of homomorphism
 $\B\phi_2$  and by $\U_{2.1}$ the kernel of homomorphism $\phi_2$.
 It is easy to see that $\U_{2.1}={\bf ker}\Psi\vert_{\U_{2.0}}$.

For every Lagrangian $L\in\U_{2.1}$, $\Pi_3(L)=0$.
 From the statement b) of the Proposition 2
it follows that one can consider the composed homomorphism

  \line{\hss$\phi_3=(\,^td_2)^{-1}\circ\Pi_2\colon \U_{2.1}
  \rightarrow \,^tE^{0.2}_2
\rightarrow K_3=
   {\Hs\over (H^1(M)\oplus H^1(\G))/H^1_{inv}(M)}$
                  \hss}
 Performing the analogous considerations for the space
  $\U_{2.0}$ we can consider the composed
  homomorphism
  $\B\phi_3=(\B{\,^td_2})^{-1}\circ\B\Pi_2\colon \U_{2.0}
  \rightarrow \,^tE^{0.2}_2
 \rightarrow K_3$.

  One can see that in this case as in the
  previous ones,
  $\B\phi_3\vert_{\U_{2.1}}=\phi_3$ and
  $\U_{3.1}={\bf ker}\Psi\vert_{\U_{3.0}}$ also,
 where we denote by $\U_{3.1}$, $\U_{3.0}$  the kernels
 of $\phi_3$ and $\B\phi_3$ correspondingly.

    For example in the case if $L$
   in (4.1) belongs to $\U_{2.1}$ then
   one can choose $\a_i$ such that $da_i=w_{i\mu}dx^\mu$
and $(\d\a)_{ij}=0$  because $\B\phi_2(L)=0$.
The equivalence class of $\a_i$ in $K_3$ is $\B\phi_3(L)$.

 In the case if $L\in \U_{3.1}$ then $\Pi_2(L)=0$.
 It means that the value of the homomorphism $\Pi_1$ (see 3.22)
on this Lagrangian is equal to the value of
this homomorphism on some $\G$-invariant Lagrangian:
  $\Pi_1(L)=\der L=\der L_{inv}$.
  From Proposition 1 it follows that
  $L=L_{inv}+w$ where closed differential $1$-form
 $w$ is defined uniquely up to closed $\G$-invariant form and
 exact form. This defines the homomorphism
 $\phi_4(L)\colon\,\U_{3.1}\rightarrow K_4=H^1(M)/(H^1_{inv(M)})_*$
where $(H^1_{inv(M)})_*$ is the image
 of $H^1_{inv(M)}$ in $H^1(M)$ under the canonical
 homomorphism.
  More formally
 $\phi_4$ can be defined as composed
homomorphism with values in the kernel
$\,^tE^{0.1}_3$ (3.21)
 of the differential $\,^td_2$
(see the statement c) of the Proposition 2):
  On the space $\U_{3.1}$ the image of $\Pi_1$ belongs to
  the image of differential $\,^td_1$ acting
  on the space $\,^tE^{0.1}_1$ in the table (3.18),
 hence
 $\phi_4=\pi\circ ({\bf id}-(\,^td_1)^{-1}\der)\colon\,
  \U_{3.1}\rightarrow E^{0.1}
 \rightarrow  K_4\subseteq \,^tE^{1.0}_3$
 where  $\pi$ is defined by (3.5).

 Analogously one can define the homomorphism
 $\B\phi_4(L)\colon\,\U_{3.0}\rightarrow K_4$.

    Similarly to previous cases
  $\B\phi_4\vert_{\U_{3.1}}=\phi_4$ and
  $\U_{4.1}={\bf ker}\Psi\vert_{\U_{4.0}}$  also,
  where we denote by $\U_{4.1}$, $\U_{4.0}$  the kernels
  of $\phi_3$ and $\B\phi_3$ correspondingly.

 From the definitions of $\phi_4$ and $\B\phi_4$ it is
 evident that Lagrangians belonging to $\U_{4.1}$
 can be reduced to $\G$-invariant by the redefinition
 on exact form (full derivative.)

  The spaces
 $\{\U_{s.1},\U_{s.0}\}$ constructed here coincide with the
spaces  $\{\V_{s.1},\V_{s.0}\}$ considered in the Introduction.
 (see (1.14, 1.15)) in the case if $\U=\L^1_1(M)$.

 These considerations can be summarized in the

   {\bf Theorem}. Let $\U$ be an arbitrary subspace
    in the space of classical mechanics Lagrangians
      for a given $[\G,M]$ pair. Let
     $\U_{0.0}$ be the subspace of $\U$
    defined by (4.1)
   which contains the weakly
   $\G$-invariant Lagrangians in $\U$.
      Then  the following relations which establish
     the classification (hierarchy) in the space $\U_{0.0}$
        are satisfied
                $$
              \matrix
                {
          &&  \U_{4.1}&\subseteq &\U_{4.0} &&&&     \cr
           && \vsubs    &    &\vsubs  &&&&               \cr
     K_4&{\buildrel \phi_4\over\longleftarrow}
             &\U_{3.1}&\subseteq &\U_{3.0} &
        {\buildrel \B\phi_4\over\longrightarrow}  &K_4&=&
                   H^1(M)\big/ (H^1_{inv}(M))_*    \cr
           && \vsubs    &    &\vsubs  &&&&                \cr
    K_3&{\buildrel \phi_3\over\longleftarrow}
          &\U_{2.1}&\subseteq &\U_{2.0} &
  {\buildrel \B\phi_3\over\longrightarrow}  &K_3&=&
               {H^1(\G,\Lambda^0(M))\over
      (H^1(M\oplus H^1(\G))\big/ H_{inv}(M)}   \cr
           && \vsubs    &    &\vsubs  &&&&            \cr
    K_2&{\buildrel \phi_2\over\longleftarrow}
        &\U_{1.1}&\subseteq &\U_{1.0}
       &{\buildrel \B\phi_2\over\longrightarrow} &K_2&=&H^2(\G)   \cr
           && \vsubs   &    &\vsubs  &&&&              \cr
    K_1&{\buildrel \phi_1\over\longleftarrow}
           &\U_{0.1}&\subseteq &\U_{0.0} &
        {\buildrel \B\phi_1\over\longrightarrow}
            &K_1&=&H^1(M)\otimes H^1(\G) \cr
      &&  &&\Psi\downarrow\phantom{\Psi} &&&&    \cr
       && &&K_0=H^1(\G) &&&&      \cr
                              }
                                          \eqno (4.2)
                             $$
 The spaces $\U_{s.\sigma}$ are intersections
  of the space $\U$ with the spaces
    $\V_{s.\sigma}$ defined in Introduction (see 1.6--1.15);
 the double filtration
 $\{\U_{s.\sigma}\}$ is subordinated to the
homomorphisms $\{\B\phi_s,\phi_s,\Psi\}$ constructed above:

\centerline {$\U_{s.0}={\bf ker}
  (\B\phi_s\colon\,\,\U_{s-1.0}\rightarrow K_s),\quad
   \U_{s.1}={\bf ker}(\phi_s\colon\,\,\U_{s-1.1}\rightarrow K_s),$}

  \centerline
   {$\U_{s.1}={\bf ker}(\Psi\colon\,\,\U_{s.0}\rightarrow K_0),\quad
        \B\phi_s\big\vert_{\U_{s-1.1}}=\phi_s$.}

 We denote the diagram (4.2) by ${\cal D}([\G,M],\U)$ and call
it the hierarchy diagram for the subspace $\U$.
In the case if $\U=\L^1_1(M)$ is the space of all
 Lagrangians of classical
 mechanics on $M$  we denote the diagram ${\cal D}([\G,M],\U)$
shortly by  ${\cal D}([\G,M])$.

  The diagram ${\cal D}([\G,M],\U)$ measures the
   differences in the spaces
  $\{\U_{s.\sigma}\}$ for an arbitrary subspace $\U$.

   We say that weakly $\G$-invariant
 Lagrangian $L\in\U$ is on the
 floor $"s"$ if $L\in\U_{s.0}$ and $L\not\in\U_{s+1.0}$.
(All Lagrangians from $\U_{4.0}$ are on the $4$-th floor.)

      We say that weakly $\G$-invariant
 Lagrangian $L$ is on the
 floor $"s_+"$ if this Lagrangian is on the  floor $"s"$ and
it belongs to $\U_{s.1}$. All other Lagrangians
 from the floor $"s"$ are on the floor $"s_-"$.

  All Lagrangians which are on the $"+"$-th floors have time
independent Noether charges, except Lagrangians in zeroth floor.

 The Lagrangians which are on the floor $"s"$ have non-trivial
image in the space $K_{s+1}$ in (4.2). The Lagrangian
 on  floor $"s_-"$ have also
non-trivial image in $K_0$ under homomorphism $\Psi$.

  Returning to the table (1.6) in Introduction
 we can conclude that a Lagrangian
which possesses the property  $"s"$ in (1.6)
and which  does not possesses the property
   $"s+1"$ in (1.6) does have non-trivial image
  in the space $K_{s+1}$.

 The evident but important corollary of the
 hierarchy diagram is that
 the floor is empty if the corresponding
  space $K_s$ is trivial.
  For example
 in the case if the first de Rham cohomology
   of configuration space are trivial then $K_1=K_4=0$
  and the zeroth floor and the  third floors are
 empty.
 In the case if the algebra $\G$ is semisimple
  only the floors $2_+,3_+,4_+$ can be non empty,
 because in this case $H^1(\G)=H^2(\G)=0$, hence
 $K_0=K_1=K_2=0$.

 The hierarchy diagram will be called trivial if all the
 spaces $K_s$ are equal to zero.

  In general the inverse statement is not valid.
 From the fact that the space $K_s$ is not trivial
 does not follow that the floor
 $"s-1"$ is not empty, because the homomorphisms in
 (4.2) are not in general surjective. For example
 homomorphism $\phi_2$ in general is not surjective
 because the map $\,^td_2$ which induces this homomorphism
 is defined on the subspace $\,^tE^{0.1}_2$
of the space $\Hs$.

 We say that the diagram ${\cal D}([\G,M],\U)$ is {\it full}
on the floor $"s_+"$ ($s<4$) if $\phi_{s+1}$ is homomorphism
 onto the space $K_s$ (surjective homomorphism),
 we say that this diagram is full on the
 floor $"s_-"$- if the restriction of $\Psi$ on $\U_{s.1}$
 is the surjective homomorphism. In the case if the diagram is full
on the floors $"s_+"$ and $"s_-"$ we say that
 it is full on the floor $"s"$.

 For a given pair $[\G,M]$ two subspaces $\U$ and $\U^\prime$
 in the space $\L^1_1(M)$ of classical mechanics
Lagrangians on $M$
 will be called equivalent with respect to the
 hierarchy if
 the images of all the
 homomorphisms $\{\phi_s,\B\phi_s,\Psi\vert_{\U_{s.1}}\}$
 for the diagram ${\cal D}([\G,M],\U^\prime)$
 coincide with the images of corresponding homomorphisms
 for the diagram  ${\cal D}([\G,M],\U)$.
 It is evident that  in this case for arbitrary $L\in\U$ there
exists $L^\prime\in\U^\prime$ such that
   $L^\prime-L$ belongs to the space $\U_{4.1}$,
  i.e.
                      $$
  L^\prime= L+L_{inv}+ {\rm full\,derivative}.
                                      \eqno (4.3)
                     $$
This construction  can be used for defining
 in the space $\U_{0.0}$ a gradation corresponding
 to the filtration (4.2) (See the examples in the next Section.)

 Now we use it for simplifying
the diagram (4.2) for physically important subspace
  $\U^{pol}$ of Lagrangians which are
 polynomial in velocities.
  Let  $\U^f=\Omega^1(M)$ be a subspace
of formal Lagrangians in $\U^{pol}$ which correspond to differential forms
by (2.4), and $\U^{sc}=\L^0(M)$ be a subspace of formal Lagrangians in
 $\U^{pol}$ which are functions on $M$.

  One can see that the space $\U^{pol}$ is equivalent to the
space $\U^{f}\oplus\U^{sc}$ with respect to the hierarchy.

 To prove it we note that every $L$ in $\U^{pol}$ can
be represented as
                 $$
   L(q,\dot q)=\sum_{n\geq 0}L_n(q,\dot q)=
           \sum_{n\geq 2}L_n(q,\dot q) +
         A_{\mu}(q)\dot q^\mu+\varphi (q)\,.
                                              \eqno (4.4)
                 $$
where  $L_n(q,\dot q)$ is the polynom of $\dot q$ of the order $n$.
Using the fact that the Lie derivative does
not change the order of polynom
  ($(\d L)_n=\d (L_n)$)
one can see that
  for  homomorphism
$\Psi$ are responsible the functions on $M$ and
 for  homomorphisms $\phi_s,\B\phi_s$  are responsible
 polynoms which are linear by velocities, i.e. differential
$1$-forms:
$\Psi(L)=\Psi(\varphi),\phi_s(L)=\B\phi_s(L)
=\phi_s(A_\mu \dot q^\mu)$. This proves the equivalence.

 The homomorphism
 $\Psi$ takes values in the subspace of $H^1(\G)$
which is isomorphic to the cohomologies of $H^1_{inv(M)}$
which are  trivial in $H^1(M)$:
 If $\delta \varphi\in H^1(\G)$ then $d\varphi\in H^1_{inv}(M)$,
      if $w\in H_{inv}^1(M)$ and $w=d\varphi$ then
 $\delta \varphi\in H^1(\G)$.

 From these facts it follows that
for the diagram ${\cal D}([\G,M],\U^{pol})$
the following additional relations are satisfied:

\line{\hss$
    \U^{pol}_{s.0}=\U^{pol}_{s.1}\oplus B,
    \quad \U^{pol}_{s.1}=\U_{4.1}\oplus A_s$.\hss      (4.5)}
  Here  $B=\U^{sc}_{0.0}/\U^{sc}_{0.1}$
  is the factorspace of functions in $\L^0(M)$
whose $\G$-symmetry variation is constant by the space
  $\L^0_{inv}(M)$ of $\G$-invariant
 functions.
   Correspondingly  $A_s=\U^{f}_{s.1}$ are the subspaces of
the space $\Omega^1(M)$ of $1$-differential forms.

   Weakly $\G$-invariant Lagrangians which belong to
the space $\U^{pol}$ differ from the
 Lagrangians in $\U^{pol}_{4.1}$ ($\G$-invariant Lagrangians
up to a full derivative) on the interaction with
"electromagnetic field" whose field strength is $\G$-invariant.
In particular a Lagrangian which is on
the floor $"s_-"$ differs from Lagrangian which is on
the floor $"s_+"$ on the interaction with "electric"
field"--$1$-form $E_\mu=\p\varphi/\p q^\mu$.
 The value of this $1$-form on every symmetries vector field
is constant: $E_\mu(q)e_i^\mu(q)=t_i$ where
 $\{e_i^\mu(q)\}$ are fundamental vector fields
 corresponding to the basis
 $\{e_i\}$ in Lie algebra
 $\G$ via the map (2.1).
 The time dependence of corresponding Noether charge
is proportional to $t_i$.

   In general for an arbitrary Lagrangian
  in $\V$ these properties
 are not satisfied. (See e.g. Example 1 in the Section 5.)

 The second physically important example of the subspace is
the subspace $\U^{dens}$ of Lagrangians on $M$ which are
 {\it densities}. (See the Remark in the 2-nd Section).

  It is easy to see that in this case
   $\U_{s.1}=\U_{s.0}$, i.e. all the floors $"s_-"$
are  empty, because the homomorphism $\Psi$ is trivial.
 (See the end of the 2-nd Section).

 We do not consider here systematically the general
 methods to handle with
calculations of the spaces $K_s$ and corresponding homomorphisms
 for an arbitrary pair $[\G,M]$, but we
 note only some points which can be useful for analyzing the
 content of the space $K_3$ in the hierarchy diagram and the groups
  $\Hs$ which generate these spaces.

  First we note that the basic example of the $[\G,M]$ pair
 is provided with the following construction.
 Let $M\subseteq N$ be the subspace of a space $N$ and the action
 of a Lie group $G$ is defined on $N$. The action of $G$ on
$N$ defines the pair $[\G,N]$ as well as the pair $[\G,M]$
where $\G=\G(G)$ is the Lie algebra
 of the group $G$.  This pair
in general cannot be generated by the action of a group
 on $M$.

    We say that the pair $[\G,M]$ is transitive if
  fundamental vector fields span the tangent bundle $TM$:
  $\forall q\in M\quad {\bf Im}\Phi\big\vert_{q}=T_qM$.
  ($\Phi$ defines the
   action of $\G$ on $M$ by (2.1).)

 For example it is the case if the action of Lie algebra
 $\G$ on $M$ is generated by transitive action of
 Lie group.

 For a given
 $[\G,M]$ we can consider the stability
subalgebra $\G_{st}(q)$ for every point $q\in M$:
 $\G_{st}(q)=\{\G\ni x\colon\,\Phi(x)\vert_{q}=0\}$.
In the case if the pair $[\G,M]$ is generated by the action
 of a group $G$,
 $\G_{st}(q)$ is isomorphic to the Lie algebra of
 stability subgroup of any point $q_0$.

  Let $[\G,M]$ be a transitive pair. (The constructions below can be
 generalized on non-transitive case also).

 If $\a$ is cocycle representing the cohomology class in
 $\Hs$ then  at arbitrary point $q_0$ it vanishes
 on the vectors
 in commutant $[\G_{st}(q_0),\G_{st}(q_0)]$.
 If this cocycle is generated by
 one-form $w$ via homomorphism $\pi$, defined by (3.5)
  ($\a=\pi w$) then
 it vanishes at  arbitrary point $q_0$ on all the vectors
  in  $\G_{st}(q_0)$. Moreover $\pi w$ is a coboundary
  iff $w$ is coboundary. Thus for any point $q\in M$
 one can consider homomorphisms
                    $$
  H^1(M){\buildrel [\pi]\over\hookrightarrow}\Hs
             {\buildrel \rho_q\over\longrightarrow}
                   H^1(\G_{st}(q_0)),\,\,[\pi]\,
         \hbox{is the injection and}\,   \rho_q\circ[\pi]=0,\,.
                                                   \eqno(4.6)
                    $$

 In the case if the pair
 $[\G,M]$ is generated by transitive action of
 Lie group $G$ (on $N\colon\, M\subseteq N$), then
  the image of the injection $[\pi]$ coincides
 with the kernel of $\rho_q$ for an arbitrary point $q$,
 because
 the homomorphisms $\rho_q$ for different points $q$ are related
 by the adjoint action of the group transformation:

   \line{\hss  $\forall (q,q_0),\,
  \forall\xi\in \G_{st}(q_0)\,\,
    \a(q,{\bf Ad}_g\xi)=  \a(q_0,\xi)\,\,{\rm if}
         \,q=g\circ q_0$.\hss      (4.7)}
 Hence in this case $K_3$ can be injected
 in the factorspace of $H^1{\G_{st}(q)}$  for any $q$:
                $$
          K_3\subseteq
      {\bf Im}\rho_q\big/{\bf Im}\rho_q\big\vert_{H^1(\G)}
                                  \eqno (4.8)
           $$
  It gives the upper estimation
for the dimension of the space $K_3$.
(To calculate (4.8) it is useful to note that from
  definition of the space $K_3$,
 (3.21) and (3.5) it follows that the
 elements of $K_3$
 are cocycles in $Z^1(\G,\L^0(M))$ factorized by cocycles
 which can be representing in the form:
   $\a=\pi w+t$, where $w$ is closed $1$-form and
   $t\in H^1(\G)$.)

  One can say more in the case if the pair $[\G,M]$
 is generated by the transitive action of the compact connected
 Lie group on the same space $M$.
 In this case taking the average
  of the group action on cocycle one comes to the
  injective homomorphism of $\Hs$ in $H^1(\G)$:
                   $$
  \d\a=0\Rightarrow
 {1\over Vol(G)}\int\a^g d\mu_G=\bar\a\colon\quad
        \Hs\hookrightarrow H^1(\G)\,.
                                                   \eqno(4.9)
                  $$
 ($d\mu_G$ is invariant measure on $G$.)

For example if the pair $[\G,M]$ is
transitive and it is generated by the action of semisimple
compact connected Lie group on the space $M$ then using
Whitehead lemmas ($H^1(\G)=H^2(\G)=0$), (4.8, 4.9)
 we see that all $K_s$ are equal to zero and
 the hierarchy diagram is trivial.

The analogous conclusions can be made too in the case if
the action is not transitive.
\bigskip

       \centerline {\bf V Examples}
\medskip
In this Section using the hierarchy
diagram (4.2) and considerations below
 we consider some examples of weakly $\G$-invariant
Lagrangians classification.

       \smallskip
                      \centerline{ Example 1}
This example is the model example. But here we describe in details
how to use the construction (4.3) for establishing gradation
corresponding to the hierarchy filtration (4.2).
              \smallskip
         We consider the following
pair $[\G,M]$. Let  $\G$ be Lie algebra $\ell_3$
     with generators $e_1,e_2,e_3$ such that
    $[e_1,e_2]=e_3, [e_2,e_3]=[e_3,e_1]=0$. Let
   a configuration space $M$ be cylinder:
          $M=\R\times S^1$ with
coordinates $(z,\varphi)$. The homomorphism $\Phi$ (see 2.1)
is defined  by the relations
                      $$
         \Phi e_1=\t e_1 ={\p\over\p z},\quad
         \Phi e_2=\t e_2 =z{\p\over\p \varphi},\quad
         \Phi e_3=\t e_3 ={\p\over\p \varphi}\,.
                                                 \eqno (5.1)
                        $$
This defines the pair $[\ell_3, S^1\times {\bf R}]$.
For this pair first we calculate the hierarchy
 diagram ${\cal D}([\ell_3,S^1\times\R])$.
We consider as $\U$ the whole
space $\L^1_1(M)$.
 From (5.1) it follows that
 every $\ell_3$-invariant Lagrangian
 has the form $F(\dot z)$ where $F$
 is an arbitrary function.

 Now  we calculate the spaces $\{K_s\}$.
 $K_0=H^1(\G)={\bf R}^2$ is generated by the cochains
       $e^1$ and  $e^2$   ($\{e^i\}$ are dual to $\{e_j\}$:
            $e^i(e_j)=\d^i_j$). The elements from $H^1(\G)$
        in components are $t_i=(a,b,0)$.
 $H^1(M)={\bf R}$ is generated by $1$-form $d\varphi$.
    Hence $K_1={\bf R^2}$ is generated by
   cochains $(d\varphi,0,0)$ and $(0,d\varphi,0)$.
  $K_2=H^2(\G)={\bf R}^2$: the cocycles
  $f_{ij}$ such that $f_{12}=0$ represent
   its cohomology class.
  It is easy to see that $H^1_{inv}(M)={\bf R}$ is generated
 by $1$-form $dz$.
  The stability subalgebra in every point
  $(z,\varphi)$ is generated by the vector $e_2-ze_3$,
 hence from (4.7) and the result for $H^1(\G)$ it follows
   that $K_3=0$.
   (The explicit calculations without (4.7) give that
 $\Hs=\R^2$ is generated by the cocycles
         $\a_i=(0,az+b,a)$;
    $d(0,az+b,a)=(0,adz,0)=\d_i ad\varphi$,
   hence  $\,^tE^{1.0}_3=\,^tE^{1.0}_2=\Hs$
  and  $K_3=0$.)

    The space $K_4={\bf R}$ is generated by the form $d\varphi$.
  We come to the result

 \line{\hss$ K_0=K_1=K_2={\bf R}^2,\quad K_3=0,
         \quad K_4={\bf R}$.            \hss(5.2)}
  We already see  that  second floor of
 ${\cal D}([ \ell_3,S^1\times {\bf R}])$  is empty.

The special analyze of the homomorphism $\phi_2$
leads to the fact that 1-st floor is empty too: the image of
 $\phi_2$ in $K_2$ is trivial
because in this special case the subspaces
 $\,^t E^{0.2}_\infty$ and  $E^{2.0}_\infty$
of $H^2(Q)$ have  zero intersection.

 Now we show that  the diagram
 ${\cal D}([ \ell_3,S^1\times {\bf R}])$ is full
on the all floors except the second one
 and study the content of the spaces
 $\{\V_{s.1},\V_{s.0}\}$.

    For this purpose we consider  the following
   $5$-dimensional  subspace of formal Lagrangians
 on $S^1\times {\bf R}$:
                              $$
    U=\{L\colon\quad L=a\dot\varphi+bz\dot\varphi
             +cz+d{\dot\varphi\over \dot z}+
           {q\over 2}{\dot\varphi^2\over\dot z}\},
                                           \eqno (5.3)
                         $$
  where $(a,b,c,d,q)$ are constants.

  We show that the diagram
   ${\cal D}([\ell_3,S^1\times\R],U)$ is full
  on all the floors except the first one. From this fact
  and from the emptiness of first floor for the diagram
 ${\cal D}([\ell_3,S^1\times\R])$ it follows that
 the whole space $\V$ of classical
 mechanics Lagrangians on $M$ is equivalent to its subspace
  $U$ with respect to the hierarchy. (See (4.3).)

  The straightforward calculations give that
 for arbitrary Lagrangian from $U$
                           $$
          \d_1 L={\cal L}_{{\p\over\p z}}L=bd\varphi+c,\,
          \d_2 L={\cal L}_{z{\p\over\p \varphi}}L=adz+bzdz+
                   d+qd\varphi,\,
          \d_3 L={\cal L}_{{\p\over\p \varphi}}L=0\,.
                                        \eqno (5.4)
                             $$
  Comparing (5,4) with (4.1) we see
 that $U=U_{0.0}$.

 Calculate the homomorphisms $\{\Psi,\phi_s,\B\phi_s\}$
 for the diagram ${\cal D}([\ell_3,S^1\times\R],U)$
  using  (5.2--5.4).
   $\phi_2=\B\phi_2=\phi_3=\B\phi_3=0$.
   $\forall L\in U,\Psi(L)=(c,d,0)\in K_0$.
 If $c=d=0$ then $L\in U_{0.1}$.
        $\phi_1(L)=\B\phi_1(L)=(bd\phi,qd\varphi,0)\in K_1$.
      If $b=q=0$ then $L\in U_{1.0}$ and if
 $c=d=b=0$ then $L\in U_{1.1}$.
 Hence $U_{3.0}=U_{2.0}=U_{1.0}$ and
correspondingly  $U_{3.1}=U_{2.1}=U_{1.0}$.
        $\phi_4(L)=\B\phi_4(L)=ad\varphi\in K_4$.
  If $a=b=q=0$ then we come to $U_{4.0}$.
  If also $c=d=0$  then we come to $U_{4.1}=0$.

 All these homomorphisms except
$\phi_2,\B\phi_2$ are surjective. Hence
 the space $\L^1_1(M)$ is reduced to its
 subspace $U$ with respect
 to the hierarchy. Moreover these homomorphisms are
injective on corresponding factor spaces.
 (${\bf Im}\Psi=U_{0.0}/U_{1.0}$,
${\bf Im}\phi_s=U_{s-1.1}/U_{s.1}$ and
${\bf Im}\B\phi_s=U_{s-1.0}/U_{s.0}$ if $s\not=2$.)

   From these considerations and (4.3) it follows that
  for every weakly $\ell_3$-invariant Lagrangian there
 exists unique Lagrangian in $U$ such that
 their difference belongs to $\V_{4.1}$:
                        $$
  \forall L\in\V_{0.0}\,\exists !(a,b,c,d,q)\colon\,
   L=F(\dot z)+\hbox{full derivative}+
              a\dot\varphi+bz\dot\varphi
             +cz+d{\dot\varphi\over \dot z}+
           {q\over 2}{\dot\varphi^2\over\dot z}\,.
                                                    \eqno (5.5)
                      $$
  Finally we come to the following gradation in
the space $\V_{0.0}$ of weakly $\ell_3$-invariant Lagrangians
on $S^1\times {\bf R}$:
                   $$
                  \matrix
                     {
   \V_{3.1}=\V_{2.1}=\V_{1.1}=\V_{4.1}\oplus
  K_4=\V_{4.1}\oplus {\bf R},\cr
  \V_{0.1}=\V_{1.1}\oplus K_1=\V_{1.1}\oplus {\bf R}^2,\,
        \V_{s.0}=\V_{s.1}\oplus K_0=\V_{s.1}\oplus {\bf R}^2\,.
                     }
                                    \eqno (5.6)
                    $$
   We consider also briefly the diagram
   ${\cal D}([ \ell_3,S^1\times \R],\U^{pol})$
   where  $\U^{pol}$ is the subspace
of Lagrangians which are polynomial in velocities.
  (See the end of 4-th Section.)
It is easy to see that $\U^{pol}$ is reduced to the
 three-dimensional space $U^{pol}$ which is the subspace
 of $U$ defined by the additional conditions $d=q=0$
 in (5.3).
The diagram ${\cal D}([ \ell_3,S^1\times \R],\U^{pol})$
is not full  on all the floors $\{s_-\}$ and on the floors
  $0_+$ and $1_+$.
 (One can show that in this case
${\bf Im}\Psi=\R\not= K_0,{\bf Im}\phi_1=\R\not= K_1$).
 The space $\U^{pol}_{0.0}$
is parametrized by three-dimensional space $U^{pol}$
(up to $\U^{pol}_{4.1}$) analogously to (5.5, 5.6)
with conditions $d=q=0$.

 We note that in (5.5) the term
$d(\dot\varphi/\dot z)$
which is responsible
 for time dependent Noether charges cannot be
considered as interaction with "electric field"
as in the case of Lagrangians in $\U^{pol}$.

  We want to note also that all the considerations
  which lead to the formula (5.6)
(except the property of homomorphism $\phi_2$)
where based on general relations which are
established by the diagram (4.2).

\smallskip
      \centerline {Example 2}
\smallskip
 Let $M=\R^n$ be an $n$-dimensional linear space
 which acts on itself by translations. It defines
  the pair $[\R^n,\R^n]$. (We identify the affine space with
corresponding linear space and with abelian algebra of translations.)
 It is easy to see that
$K_0=\R^n,K_2=\R^n\wedge\R^n,K_1=K_3=K_4=0$. The space
  of Lagrangians on $\R^n$ is equivalent to the
 space $U=\{L\colon\, L=w_2(q,\dot q)+w_1(q)\}$
with respect to the hierarchy,
where $w_2$, $w_1$ are $2$-cocycle and $1$-cocycle
 correspondingly on the Lie algebra $\R^n$. (The tangent vectors
 on  $\R^n$ can be identified with points.) In the same way like in
 (5.3--5.5) we come to the statement that every weakly
 $\G$-invariant Lagrangian in this case has the form

 \line{\hss $L=F(\dot q^1,\dots,\dot q^n)+$ full derivative
           $+B_{ik}q^i \dot q^k+E_i q^i$.\hss}
It describes the interaction  with constant
"magnetic" and "electric" fields.
 (Compare with (1.5)).  The corresponding Noether charges are
   $N_i(q,\dot q,t)=\p L/\p\dot q^i-B_{ik}q^k-E_i t.$
  The corresponding gradation of the space $\V_{0.0}$
is the following:

\line{\hss$\V_{4.1}=\V_{3.1}=\V_{2.1},\V_{1.1}= \V_{0.1}=
        \V_{4.1}\oplus \R^{{n(n-1)\over 2}},
          \V_{s.0}=\V_{s.1}\oplus \R^n$.\hss}
   This case is famous in literature as "arising of constant
 magnetic filed as central extension of translations
  algebra [\Jack]."
 \smallskip
         \centerline {Example 3. $so(3)$ algebra.}
     \smallskip
 In this example we consider the Lie algebra $so(3)$
which is the special case of semisimple algebra.
  Let $M=\R^3$ be $3$-dimensional linear
 space with cartesian coordinates
  $(x^1,x^2,x^3)$.  We consider first
  the  pairs $[so(3),\R^3]$ and $[so(3),S^2]$
 where $S^2$ is the sphere $x^i x^i=1$ in $\R^3$ and the action
 of $so(3)$ on $\R^3$ is generated by the standard action
 of the group $SO(3)$ on $\R^3$:
 if $\{e_1,e_2,e_3\}$ is a basis in
  $so(3)$ such that  $[e_i,e_j]=\vare_{ijk}e_k$ then
  $\Phi(e_i)=\t L_i=-\vare_{ijk}x^j\p/\p x^k$.
 For the pair $[so(3),\R^3]$  the hierarchy diagram is trivial
because $SO(3)$ is semisimple compact group.
 (See the end of the Section IV.)

 Alternatively one can see it by the following
 explicit calculations:
 From commutation relations it is evident
 that $H^1(so(3))=H^2(so(3))=0$.
 Hence $K_0=K_1=K_2=K_4=0$.
 If $\a_i$ is a cocycle with values in functions
 on $S^2$ then
 $0=\d\a=\t L_i\a_k-\t L_k\a_i-\varepsilon_{ijk}\a_k$.
 Hence $\t L^2\a_k=\t L_k(\t L_i\a_i)=\t L_k F$
      and
 $\a_k=\d \t F$ is coboundary
 where $\t F=\sum_l{F^l\over l(l+1)}$.
  ($F^l$ is defined by the expansion over the
spherical harmonics of $F$.
 The term $F^0=0$ because it leads to cocycle
 in constants and $H^1(so(3))=0$.)
 Hence $K_3=0$ also.

 The calculations and the result are the same for the
diagram $[so(3),\R^3]$.

  We come to the result that all weakly $so(3)$
 invariant Lagrangians of classical mechanics on  $R^3$  and
  on $S^2$ are exhausted by $so(3)$-invariant ones
 (up to a full derivative).

  Now bearing in mind the construction (4.6)
 we modify little bit this example considering
 instead the sphere
 $S^2$ the domain in it, the sphere without North pole
 (punctured sphere)
  $S^2\min N]$ ($x^3\not=1$). Thus we come from the pair
  $[so(3),S^2]$ to the pair $[so(3),S^2\min N]$.
In the same way we come to the pair $[so(3),\R^3\min L_+]$,
taking out the ray $L_+$ ($x^1=0,x^2=0,x^3\geq 0$) from $\R^3$.

 The essential difference of these pairs from the previous ones
 is that they cannot be generated by the action of the corresponding
 Lie group.

 We perform the calculations
 for the diagram ${\cal D}([so(3),S^2\min N])$.

 It is evident that for this diagram
  $K_0=K_1=K_2=K_4=0$, also.
  Now we show that for this diagram
 $K_3=\R$ and this hierarchy is full.

 The stability algebra for this pair is one dimensional, hence
  from (4.6--4.8) it follows that $K_3=0$ or $K_3=\R$.
 It remains to prove that
  $K_3$ is not trivial.

 To show it we consider the Lagrangian
  $L$ which corresponds to the differential form
 $A=-(1+cos\theta)d\varphi$ on the punctured sphere $S^2\min N$.
 ($\theta,\varphi$ are spherical coordinates.)
  The two-form $dA=F=sin\theta d\theta d\varphi$ corresponding
to its motion equations is $so(3)$-invariant, hence
  this Lagrangian is weakly $so(3)$-invariant.
  On the other hand it cannot be reduced to
$so(3)$-invariant by redefinition on a full derivative
 $df$ because $so(3)$-invariant $1$-form
 on the sphere is equal to zero. Hence,
 because all other spaces $K_s$
 are equal to zero.
this Lagrangian belongs to the floor $2_+$.
 We come to the result:
                  $$
      K_3=H^1(so(3),\L^0(S^2\min N))=\R\quad{\rm and}\,\,
  \phi_2(A)\in K_3\not=0,\,
                                       \eqno(5.7)
                 $$
 For this special case the explicit realization of
    (4.6-4.8) is the following:
 We identify the vectors in $\R^3$ with the vectors
 in the linear space
 of the Lie algebra $so(3)$ by the linear map
 $\gamma\colon\,(x^1,x^2,x^3)\rightarrow x^1e_1+x^2e_2+x^3e_3$.
  For any point $x\in S^2$ the corresponding stability
   subalgebra is generated by $\gamma(x)$.
   To (4.6--4.7) corresponds the following statement:
   If $\a$ is a $1$-cocycle with values in functions on
   the punctured sphere then

\line{\hss$\a(x,\gamma(x))=x^i\a_i(x)$ is a constant on
 the sphere,\hss    (5.8a)}
\line{\hss this constant is equal to zero iff this
 cocycle is a coboundary.\hss    (5.8b)}

  (This statement can be easily proved
  in a straightforward way without using (4.6,4.7)).

We proved that $K_3=\R$ and all other $K_s$ are equal to zero
 and presented in (5.7) the Lagrangian with non-trivial
 image in $K_3$. Hence
 the hierarchy diagram
 ${\cal D}([so(3),S^2\min N])$ is full on all the floors
 and the space of classiclal
 mechanics Lagrangians is equivalent to the one-dimensional space
 $U=\{L\colon\, L=-q(1+cos\theta\dot\varphi)\}$
 with respect to this hierarchy. So using (4.3) we arrive
 at the statement that
 every weakly $so(3)$ invariant Lagrangian on
the punctured sphere has the form
                        $$
               L=L_{inv}+{\rm full\, derivatives}-
                g(1+cos\theta)\dot\varphi\,.
                                                     \eqno (5.9)
                      $$
 In the case $g\not=0$ it belongs to the
 floor $2_+$ of the hierarchy.

 The calculations for the diagram
${\cal D}[so(3),\R^3\min l_+]$ are analogous and
the result is the same:
 every weakly $so(3)$ invariant Lagrangian on
the  $\R^3\min l_+$ has the form  (5.9).

 One can see that in the case
 if $L_{inv}$ is free particle Lagrangian, then
 (5.9) corresponds to the
 Lagrangian which describes
 the interaction of particle with Dirac monopole.

 The explicit calculations for (5.7)
 give that $\phi_2(L)$ for the Lagrangian (5.9)
 is equal to the cohomology class in $H^1(so(3),S^2\min N)$
 of the following cocycle:
                              $$
 \a_1= -gctg{\theta\over 2}cos\varphi,\,
 \a_2= -gctg{\theta\over 2}sin\varphi,\,
                  \a_3=g,\quad
                 (\d L=d\a,\,\d\a=0)
                                               \eqno (5.10)
                              $$
 and $\a_ix^i=-g$.

  Finally we make the following remark about the Lagrangian (5.9)

   Via stereographic projection of the
 punctured sphere on $\R^2$
 one comes from the pair $[so(3],S^2\min N]$ to the pair
 $[so(3),\R^2]$, where the fundamental vector
field  corresponding to $e_3$ corresponds to rotations and
 fundamental vector
 fields  corresponding to $e_1,e_2$ correspond to
 non-linear infinitesimal transformations.
 The weakly $so(3)$-invariant Lagrangian (5.9) transforms to the
                         $$
      L={m(\dot u^2+\dot v^2)\over 2(1+u^2+v^2)^2}+
              g{u\dot v-v\dot u\over 1+u^2+v^2}
                                                   \eqno (5.11)
                        $$
in the case if $L_{inv}$ is free particle Lagrangian.

  The Lagrangian (5.11) in the case $g=0$ is
 strictly related with the Lagrangian
 which describes the interaction
 of free particle in $2$-dimensional plane with
 Coulomb potential.
 To the vector fields $\t e_1,\t e_2$ correspond
 so called hiiden symmetries of Coulomb interaction
which lead to Runge--Lentz vector [\Stern].
 So, Lagrangian (5.11) leads to the Lagrangian
 which possesses essentially generalized hidden
 symmetries of two-dimensional Coulomb potential.
 These consideration deal with
 so called higher symmetries  which
 are not in the frame of this paper.

           \smallskip
     \centerline {Example 4. Galilean and Poincar\'e Lie algebras}
\smallskip

   To threat these algebras simultaneously we consider
 $1$-parametric family of Poincar\'e Lie algebras $\G(\P_c)$
($c$ is the "velocity of light"). Their action (2.1) on the space
 $\R^4$ with cartesian coordinates $(t,x^1,x^2,x^3)$
is generated in a standard way via the following
fundamental vector fields:
                        $$
        \t p_0={\p\over\p t},\,
      \t p_i={\p\over\p x^i},\,
      \t B_i=t{\p\over\p x^i}+{1\over c^2}x^i{\p\over\p t},\,
      \t L_i=-\vare_{ijk}x^j{\p\over\p x^k}\,.
                                                      \eqno (5.12)
                       $$
which correspond to its basis. The relations (5.12) define
 the pair $[\G(\P_c),\R^4]$.

In the case  $c\rightarrow\infty$ Lie algebra
$\G({\cal P}_c)$ is contracted to the Lie
algebra $\G({\Ga})$ of Galilean group
(non-relativistic limit) which we denote  also by
 $\G({\cal P}_\infty)$. (All the  commutation relations
 of basis vectors in $\G(\P_c)$ do not depend on $c$, except the
relations
  $[\t B_i,\t B_k]=-1/c^2\vare_{ijk}\t L_k,
[\t p_i,\t B_k]=-1/c^2 p_0\d_{ik}$ which tend to zero
if $c$ tends to zero.)

Correspondingly to (5.12) the action of the Galilean Lie
algebra  $\G(\P_\infty)$ on $\R^4$ is generated via the vector fields:
                        $$
        {\p\over\p t},\quad
      {\p\over\p x^i},\quad
      t{\p\over\p x^i},\quad
      L_i=-\vare_{ijk}x^j{\p\over\p x^k}\,.
                                                      \eqno (5.13)
                       $$
 It defines the pair $[\G(\P_\infty),\R^4]$.
 (The vector field corresponding to Lorentz boost transforms
to vector field corresponding to special Galilean transformation.)

 The first two cohomologies groups for algebras $\G(\P_c)$ are
                       $$
                     \matrix
                           {
          H^1(\G(\P_c))=0,
   &H^2(\G(\P_c))=0,  &\,\,{\rm if}\,\, c\not=\infty\cr
    H^1(\G(\P_\infty))=0,&
      H^2(\G(\P_\infty))=\R.& \cr
                       }
                                            \eqno (5.14)
                       $$
   The second cohomology group of the Galilean
 Lie algebra is generated by
 $2$-cocycle $c_B$ (Bargmann cocycle) whose
 non-vanishing components in the basis (5.13)
are only

\line{\hss $c_B(p_i,B_j)=-c_B(B_j,p_i)=\d_{ij}$.\hss    (5.15)}

 The relations (5.14) make trivial the calculations of the all
 spaces  $K_s$ except the space $K_3$ for the hierarchy diagram
 ${\cal D}(\G(\P_c),\R^4)$.  From the formula (4.6)
it follows that $K_3=0$ because
the stability subalgebra of every point
 $(t_0,x^i_0)$ in $\R^4$ is isomorphic to the
 subalgebra generated
 by the vectors $(L_i,B_j)$ which has only
 trivial $1$-cocycles.

Hence for the hierarchy diagram ${\cal D}(\G(\P_c),\R^4)$
all the spaces $K_s$ are equal to zero,
 except $K_2$ which is equal to $\R$ for Galilean algebra and
which is equal to zero for Poincar\'e algebra.

 We see that the hierarchy diagram
 ${\cal D}(\G(\P_c),\R^4)$ for Poincar\'e algebra
 is trivial.
 For Galilean algebra in the diagram
${\cal D}(\G(\P_\infty),\R^4)$ only the floors $1_+,4_+$
can be non-empty.

 It has to be noted
 that the space of Lagrangians $L(t,x^i,dt/d\tau, dx^i/d\tau)$
 in $\R^4$ is more
wide that the space of classical mechanics
Lagrangians $L(x^i, dx^i/dt)$ on the configuration space $\R^3$.
 To every  Lagrangian in $\R^3$ according to (2.16)
corresponds Lagrangian
 which is a density  in $\R^4$.
On the other hand to every Lagrangian--density  $L$ in $\R^4$
 which does not depend explicitly
on time corresponds the classical mechanics Lagrangian,
if we put the parameter $\tau=t$.
 For example to the Lagrangian
 of free non-relativistic particle
 corresponds the density in $\R^4$
  $L_{nonrel}=m\dot x^i\dot x^i/ 2\dot t$
  and to the Lagrangian of free relativistic particle
 corresponds the density
 $L_{rel}(c)=-mc\sqrt{c^2\dot t^2-\dot x^i\dot x^i}$.
  (The $\dot x^i,\dot t$ means  derivatives of $x,t$ with
 respect to the parameter $\tau$.)
 The Lagrangian $L_{rel}(c)+mc^2\dot t$ which differs from
 $L_{rel}(c)$ on the full derivative
 tends to $L_{nonrel}$ if $c\rightarrow \infty$.

 The Lagrangian $L_{rel}(c)$ is $\G(\P_c)$-invariant.
 $L_{rel}(c)$ is the unique (up to multiplier)
$\G(\P_c)$-invariant Lagrangian in the space of densities
 on $\R^4$.

 On the other hand
  it is evident that there are no $\G(\P_\infty)$-invariant
Lagrangians in the space of densities on $\R^4$, except
 trivial ones.
 The Lagrangian $L_{nonrel}$
 is weakly $\G(\P_\infty)$-invariant:
                          $$
         {\cal L}_{p_0}L_{nonrel}=
           {\cal L}_{p_i}L_{nonrel}=
            {\cal L}_{L_i}L_{nonrel}=0,\,
          {\cal L}_{B_i}L_{nonrel}=m\dot x^i\,.
                                            \eqno (5.16)
                         $$
  From  (5.16) it follows that $\d L=d\a$ where
  the  values of the cocycle $\a$ on the
basis vectors $B_i$  are equal to $mx^i$,
 on all other basis vectors $\a$ is equal to zero.
Hence

\line
{\hss$\d\a=mc_B,$\hss  (5.17)}
 where $c_B$ is Bargmann cocycle (5.15).

 We proved before that for the diagram
 ${\cal D}([\G(\P_\infty)],\R^4)$, $K_2=\R$ and all other
 $K_s$ are equal to zero.  On the other hand
 it follows from (5.16,5.17)
  that the homomorphism $\phi_2$ for the diagram
 ${\cal D}([\G(\P_\infty)],\R^4)$ has non-trivial image in $K_2$
  at the Lagrangian
  $L_{nonrel}$.
 Hence the space of all weakly
 $\G(\P_\infty)$-invariant Lagrangians
 on $\R^4$ is equivalent to the one-dimensional space generated
  by the Lagrangian of free non-relativistic particle.

  We come to the following conclusion:

 Every   weakly $\G(\P_\infty)$-invariant Lagrangian
 which is a density in $\R^4$
 belongs to the floor $1_+$ and
 is proportional to $L_{nonrel}$
 (up to full derivative and a constant).
 The floor $4_+$ contains only trivial Lagrangians.

 Correspondingly for the Poincar\'e algebra
 every weakly $\G(\P_c)$-invariant Lagrangian-density
 coincides (up to full derivatives)
 with   $\G(\P_c)$-invariant Lagrangian $L_{rel}(c)$.

 On one hand to the contraction of Poincar\'e
 algebra to Galilean algebra
 corresponds  the arising of Bargmann cocycle.
 On the other hand the unique non-trivial component
 $\V_{4.1}$ of the hierarchy diagram
 for Poincar\'e algebra transforms
 to the unique non-trivial component $\V_{1.1}$ of
 the hierarchy diagram for Galilean algebra.

 The vanishing of  $ H^2(\G({\cal P}_c))$ is the reason
 why in relativistic quantum mechanics the projective
 representation of Poincar\'e symmetries in the space
 of states (which are rays in linear Hilbert space)
 can be reduced to
 linear one and because of (5.15) it is not the case in
 non-relativistic mechanics.
 The considerations of this example  are
 reflection of this phenomenon.
\bigskip
       \centerline {\bf VI Discussions}
 \smallskip
The problem considered here and the technique which we used
to study it can be generalized in a few directions.
 The considerations of this paper can be easily translated
 to Hamiltonian language.
 One can consider the classification of Lagrangians not only
for symmetries induced by point transformations of
 configuration space but by the so called higher symmetries.
For example from this point of view it is interesting to analyze
 the generalized Runge--Lentz symmetries (see the end of
 Example 3 in Section V).

 It is interesting to apply this method
to supersymmetrical case [\Ann]. It seems to be interesting
 also to analyze the phenomena of spin-like
 transformations (1.9) arising for Lagrangians
 from the second floor of the hierarchy (4.2),
 in order to apply it to Dirac monopoles [\Ners].

 We hope that a generalization of this method
on field theory Lagrangians will be fruitful.
 From this point of view we want to note the relations
 of our considerations with the problem of the
 Ward identities anomaly absence
 for field theory Lagrangians
 which possess classically  the given symmetry [\Stor,\Pig].

 To develop this technique for
 field theory Lagrangians,
 the first order formalism and multisymplectic
 formalism become very useful [\Stern].
 We would wish to develop these considerations on the
 firm ground of investigations of A.M. Vinogradov and his
 collaborators [\Vin].

 On the other hand, in our opinion, the method
 considered in this paper
 maybe is more important than the problem we applied it to.

  We give here only three examples, one of them pure mathematical,
 where the calculations of double
comlex cohomology (the method we use in this paper)
 make a bridge between the corresponding structures.

 1. {\it The calculation of de Rham cohomology in terms of
 Chech cohomology}.

  When manifold $M$ is covered by the family
$\{U_\a\}$ of open sets one can consider Chech cohomology of this covering.
Then one can consider double complex of
 $q$-forms which are defined on the sets $\{U_\a\}$. The differential
 $Q$ of this complex is the sum of de Rham exterior
 differential and Chech
 differential. Considering the differential $Q$
 "perturbatively" around Chech differential one arrives naturally
 at de Rham cohomology of $M$, hence
 the "perturbative" calculations around de Rham differential lead
 in general to calculation of spectral sequence which tends
 to de Rham cohomology of $M$.
 In the case if the covering is Leray covering, i.e.
 all the sets and their intersections are convex connected
 sets then Chech cohomology coincide with de Rham one;
 the application of Poincar\'e lemma
 reduces spectral sequence calculations to trivial
 resolutions of descent equations. But practically it is
 more convenient to use for calculations
 a suitable covering which generally
 is not a Leray covering.
 (See for details e.g. [\Postn].).

 2. {\it The relations between Hamiltonian reduction method and
 BRST cohomology for classical mechanics}

  One can say that the relations between these two methods
are encoded in the cohomology of double complex differential
$Q=\p+\d$ in the case if constraints form Lie algebra (so called
 closed groups.) Here $\p$ corresponds to Koszul differential of
complex generated by constraints and $\d$ is differential
 corresponding to Hamiltonian vector fields
 which are induced by these constraints. Perturbative expansion
 of $Q$ around $\d$ leads to standard Hamiltonian methods,
 and expansion around $\p$ leads to BRST. In the case if
 constraints form so called open group, one has to consider
  the corresponding filtered space instead of this double complex
 [\Kost,\Dub,\Hen]. This approach seems to be very
 fruitful.

 3. {\it Local BRST Cohomology}

 Considering BRST physical observables as integrals of
local functions one comes naturally to differential
 $Q=s+d$, where $s$ is BRST differential,
 acting on integrand which is
 local function and $d$ is the usual de Rham differential.
 It turns out that the consideration of
 cohomology of this double complex is a very powerful tool
 for BRST cohomology investigations in field theory, especially
 in Lagrangian framework.
 (See [\Dix,\Barn,\Pig,\Sta] and citations there).

 In spite of these examples one has to note that the method
 of spectral sequences was not used actively in these calculations.

 May be at first the method of spectral sequences was
 applied in physics
 by J. Dixon in [\Dix] in analysis of local BRST cohomology.
 In series of works the so called
 method of descent equations which is in fact
 a special case reminiscent of this technique was
 applied successfully to these problems.
 (See the review [\Pig] and the papers citated there).
 Nowadays the technique of spectral sequences seems to
 be not very popular in theoretical physics.
 We hope to pay attention to importance
 of this technique which
 is used here in a simple physical frame.
 In principal using the method
 "Deus ex machina" one can formulate the hierarchy without
 using explicitly the method  used in this paper which indeed
 seems to be very tedious. But in our opinion  this method
 is inherent to this problem and it is the
 adequate technique in other important problems
 such as constrained dynamics theory; it may have
  useful applications in future.
     \bigskip

\centerline {\bf VII Acknowledgments}
 The work was supported in part by Armenian
 National Science Foundation and by INTAS-RFBR
 Grant No 95-0829.

 We are grateful to G.Barnich, F.Brandt, R.L. Mkrtchyan and
A.P.Nersessian  for  fruitful discussions
and important remarks.

 The creative atmosphere of the conference
"Secondary Calculus and Comological Physics"
which was organized by A.M. Vinogradov's school
gave the strong impulse for this work.
We are very grateful to
 all participants of this conference and especially
to N. Maggiore, M. Castrillon and J. Stasheff who
encouraged to finish this work.

Finally one of us (O.M.) wants to use opportunity to express
his deep gratitude to O.Piguet whose indirect influence on
this paper was extremely high.

     \bigskip
            \centerline {\bf Appendix 1.
        Lie algebra cohomologies}
              \bigskip

 Let $\G$ be Lie algebra and $A$ be a linear space
which is module on $\G$,
i.e. the action of $\G$ on $A$
which respects the structure of the Lie algebra $\G$
and the space $A$ is defined:
                          $$
                    \eqalign
                        {
    h\in \G,m\in A\quad (h,m)&\rightarrow h\circ m\in A\colon\cr
    (\lambda h_1+\mu h_2)\circ m&=
         \lambda (h_1\circ m) + \mu (h_2\circ m),\quad
         (\lambda,\mu\in {\bf R}) \cr
          h\circ (\lambda m_1+\mu m_2)&=
                 \lambda (h\circ m_1)       +
                  \mu (h\circ m_2), \cr
         h_1\circ(h_2\circ m)- h_2\circ(h_1\circ m)&=
                   [h_1,h_2]\circ m\,.\cr
                         }
                                           \eqno (A1.1)
                           $$
  ($[\,,\,]$ defines commutator in $\G$. $A$
and $\G$ are linear spaces on ${\bf R}$).

 The complex $(C^q(\G,A),\d)$ of cochains
 can be defined in the following way.
 Let $C^q(\G,A)$ be a space of
skewsymmetric $q$-linear functions on $\G$ ($q$-cochains)
 which take values in $A$ (If $q=0$, $C^0(\G,A)=A$).
 $\G$-differential $\d$ on $\{C^q\}$
 $\d\colon C^q\rightarrow C^{q+1},\,\, \d^2=0$
   is defined in the following way:
                      $$
                   \eqalign
                      {
         \d\colon C^0\rightarrow C^1\quad
       &(\d c)(h)=h\circ c,(c\in C^0=A)\cr
  \d\colon C^1\rightarrow C^2\quad
   &(\d c)(h_1,h_2)=
     h_1\circ c(h_2)-h_2\circ c(h_1)-c([h_1,h_2]),\cr
                         }
                                            \eqno(A1.2)
                        $$
 and so on:
                      $$
                    \eqalign
                         {
      \d\colon C^q\rightarrow C^{q+1}\quad
          (\d c)(h_1,\dots,h_{q+1})=
          \sum_{1\leq i\leq q+1}
                  (-1)^{i+1}
     &h_i\circ c(h_1,\dots,{\hat h_i}\dots,h_{q+1})+\cr
              \sum_{1\leq i<j \leq q+1}
                     (-1)^{i+j}
      c([&h_i,h_j],h_1,\dots,{\hat h_i},\dots,
        {\hat h_j}\dots,h_{q+1})\cr
                        }
                       $$
  ( ${\hat h_i}$ means omitting of the variable $h_i$)).
The cohomologies $H^q(\G,A)$ of the complex $(\{C^q\},\d)$ are called
 cohomologies of Lie algebra $\G$ with coefficients
in the module $A$.
(See in details for example [\Postn].)
                        $$
        H^q(\G,A)=
   \left({\bf ker}\,\d\colon C^q\rightarrow C^{q+1}\right)
                   \big/
         \left({\bf Im}\,\d\colon C^{q-1}\rightarrow C^q\right)\,.
                       $$

 If module A is ${\bf R}$
 and $\G$ acts trivially on it: $h\circ \lambda=0$,
 $C^q(\G,{\bf R})$ is denoted by $C^q(\G)$ and correspondingly
 $H^q(\G,{\bf R})$ is denoted by $H^q(\G)$.
 In this case cochains are constant antisymmetrical tensors
 and $\G$-differential $\d$ is expressed only via
 structure constants $\{t_{ik}^n\}$ of Lie algebra $\G$.

 $H^0(\G)={\bf R}$, $H^1(\G)$ is defined
 by the solutions of the equation
    $c_{ik}^m b_m=0$ and it is nothing but the space dual to
    the $\G/[\G,\G]$.

 In a case if $\G$ is abelian
 $H^q(\G)=C^q(\G)=(\wedge \G^*)^q$ where
 $\G^*$ is the linear space dual to the linear space of $\G$.

 In a case if $\G$ is semisimple Lie algebra then
 $H^1{\G}$=$H^2{\G}$=0.
 This statement is valid in a general case too.
 Very important Whitehead lemmas state that
 if $\G$ is semisimple Lie algebra then
  $H^1(\G,A)= H^2(\G,A)=0$
 in the case if $A$ is an arbitrary module which
is {\it finite-dimensional} vector space on ${\bf R}$ [\Postn]

         \bigskip

     \centerline {\bf Appendix 2.
  Double complex and its spectral sequences.}
                  \medskip
 Now we give a brief sketch on the topic how
 to apply spectral sequences technique
for calculations of cohomology of double complexes.
 (See for the details for example [\Postn].)

Let $E^{**}=\{E^{p.q}\}$ $(p,q=0,1,2,...)$ be a
 family of abelian groups
(modules, vector spaces) on which are defined two differentials
$\p_1$ and $\p_2$ which define complexes in rows and in columns of
 $E^{*.*}$ and which commute with each other:
                            $$
          \p_1\colon E^{p.q}\rightarrow E^{p.q+1}\,,\p_1^2=0,\,
          \p_2\colon E^{p.q}\rightarrow E^{p+1.q}\,,\p_2^2=0,\,
                    \p_1\p_2=\p_2\p_1\,.
                                       \eqno (A2.1)
                            $$
   $\{E^{**},\p_1,\p_2\}$ is called double complex.

  ( It is convenient to consider $E^{p.q}$ for all
integers $p$ and $q$ fixing that $E^{p.q}=0$ if $p<0$ or $q<0$.)

 One can consider  "antidiagonals":
   ${\cal D}^m=\{E^{p.m-p}\}$ ($p=0,1,...,m$)
  which form complex with differential
                             $$
                        Q=(-1)^q\p_2+\p_1
                                                  \eqno (A2.2)
                             $$
 which evidently obeys to condition $Q^2=0$.
                               $$
       0\rightarrow{\cal D}^0{\buildrel Q\over\rightarrow}
          {\cal D}^1{\buildrel Q\over\rightarrow}{\cal D}^2
                   \rightarrow\dots\,.
                                                 \eqno (A2.3)
                               $$
            The cohomologies $H^m(Q)$ of this complex
are called  the cohomologies of double complex
$(E^{**},\p_1,\p_2)$.

 The rows and the columns complexes define the cohomologies
  $H(\p_1)$ and  $H(\p_2)$ of $E^{**}$.

One can consider the filtration corresponding to the
double complex  $\{E^{*.*},\p_1,\p_2\}$
                          $$
    \dots\subseteq X^m\subseteq
  X^{m+1}\subseteq\dots\subseteq X^1\subseteq X^0
                                              \eqno (A2.4)
                           $$

                     $$
  {\rm where}\qquad\quad
     X^k= \bigoplus_{q\geq 0,p\geq k} E^{p.q}
                                      \eqno (A2.5)
                    $$

 and sequence of the spaces  $\{E^{p.q}_r\}$
 $(r=0,1,2,\dots$ corresponding to this filtration
                           $$
             E^{p.q}_r =Z^{p.q}_r\big/B^{p.q}_r\quad
                  (E^{p.q}_0=E^{p.q})\,.
                                                \eqno(A2.6)
                             $$
In (A2.6) $Z^{p.q}_r$ ("$r$-th order cocycles")
 is the space of the elements in $E^{p.q}$
 which are leader terms of
 cocycles of the differential $Q$ up to $r$--th order
w.r.t. the filtration (A2.4), i.e.
                          $$
       \{Z^{p.q}_r\}=\{E^{p.q}_r\ni c\colon\quad
       \exists {\t c}= c
           (mod\, X_{p+1}) \,{\rm such\,that}\,
             Q{\t c}=0(mod\,X_{p+r})\}\,.
                                                \eqno(A2.7)
                          $$
   It means that there exists ${\t c}=(c,c_1,c_2,\dots,c_{r-1})$
where $c_i\in E^{p+i.q-i}$  such that

\noindent$Q (c,c_1,c_2,\dots,c_{r-1})\subseteq X_{p+r}\,$:
                          $$
     \p_1 c=0,\p_2 c=\p_1 c_1,\p_2 c_1=\p_1 c_2,\dots,
    \p_2 c_{r-2}=\p_1 c_{r-1},\,{\rm so}\,
     Q {\t c}=\p_2 c_{r-1}\in X_{p+r}\,.
                           $$
 Correspondingly $B^{p.q}_r$ is the space
of up to $r$--th order borders:
                          $$
       \{B^{p.q}_r\}=\{E^{p.q}_r\ni c\colon\quad
       {\cal 9} {\t b}\in X_{p-r+1}\,{\rm such}\quad{\rm that}\,
             Q{\t b}=c\,.
                                                \eqno(A2.8)
                          $$
   It means that there exist ${\t c}=(b_0,b_1,b_2,\dots,b_{r-1})$
where $b_i\in E^{p-i.q+i}$  and

\noindent $ Q (b_0,b_1,b_2,\dots,b_{r-1})=c$:
                          $$
     \p_1 b_0+\p_2 b_1=c,\p_1 b_1+\p_2 b_2=0,
       \p_1 b_2+\p_2 b_3=0,\dots,
                \p_1 b_{r-1}=0\,.
                                 \eqno (A2.9)
                           $$
  For example $E^{p.q}_1=H(\p_1,E^{p.q})$.

 We denote by $[c]_r$ the equivalence class of the
 element $c$ in the $E^{p.q}_r$ if $c\in Z^{p.q}_r$.

 It is easy to see that the sequence $\{E^{p.q}_r\}$
   $r=0,1,2,\dots$ is stabilized after finite number of the steps:
($E^{p.q}_{r_0}=E^{p.q}_{r_0+1}=\dots=E^{p.q}_\infty$,
 where $r_0=max\{p+1,q+1\}$.

  Let $H^m(Q, X_p)$
 be cohomologies groups of double complex truncated by
 filtration (A2.4) (we come to $H^m(Q, X_p)$   considering
  $\{{\cal D}\cap X^p,Q\}$
 as subcomplex of (A2.3), $H^m(Q)=H^m(Q,X^0)$.
 We denote by  $_{(p)}H^m(Q)$ the image of
$H^m(Q, X_p)$ in $H(Q)$   under
 the  homomorphism
 induced by the embedding ${\cal D}\cup X_p\rightarrow {\cal D}$.
  The spaces $_{(p)}H^m(Q)$    are embedded in each other
                      $$
   0\subseteq \,_{(m)}H^m(Q)\subseteq \,_{(m-1)}H^m(Q)\subseteq
        \dots\,_{(1)}H^m(Q)\subseteq
                 \,_{(0)}H^m(Q)=H^m(Q)\,.
                                                \eqno(A2.10)
                     $$

 The spaces $E^{p.q.}_\infty$ considered above
 are related with (A2.10) by
 the following relations:
                              $$
         E^{p.m-p}_\infty=_{(p)}H^m(Q)
                         \big/
                    _{(p+1)}H^m(Q)   \,.
                                             \eqno (A2.11)
                               $$
 In particular  $E^{0.m}_\infty$ is canonically embedded
 in $H^m(Q)$.

  The formula (A2.11) is the basic formula which expresses
 the cohomology
$H(Q)$ of the double complex $\{E^{p.q},\p_1,\p_2\}$
in terms of $\{E^{p.q}_\infty\}$.
 From (A2.10, A2.11) it follows that
                             $$
           H^m(Q)\simeq\bigoplus_{i=0}^{m} E^{p-i.i}\,.
                                                \eqno (A2.12)
                            $$
The essential difference of (A2.12) from  (A2.11) is that
 in (A2.12) the isomorphism of l.h.s. and of r.h.s.
{\it is not canonical}.

 The importance of the sequence $\{E^{*.*}_r\}$ ($r=0,1,2,\dots$)
is explained by the fact that its terms
(and so $\{E^{*.*}_\infty\}$) can be calculated in a recurrent way.
Namely one can consider differentials (See for details [\Postn.])
  $d_r\colon\, E^{p.q}_r\rightarrow E^{p+r.q+1-r}_r$
 such that  $\{E^{*.*}_r, d_r\}$ form spectral sequence, i.e.
                            $$
              E^{*.*}_{r+1}=H(d_r,E^{*.*}_r).
                                           \eqno (A2.13)
                            $$
The differentials $d_r$ are constructed in the following way:
   $d_0=\p_1\colon\, E^{p.q}=E^{p.q}_0
\rightarrow E^{p.q+1}=E^{p.q+1}_0$.

   \noindent
If $c\in E^{p.q}$ and $\p_1 c=0\leftrightarrow
 [c]_1\in E^{p.q}_1$ then
 $d_1[c]=[\p_2c],\,d_1\colon\,E^{p.q}_1
      \rightarrow E^{p+1.q}_1$.

\noindent  In general case for $[c]_r\in E^{p.q}_r$
  $d_r[c]_r=[Q{\t c}]_r\,\,d_r\colon\,
 E^{p.q}_r\rightarrow E^{p+r.q+1-r}_1$,

 where $\t c\colon\, c-\t c\in X^{p+r}$ (see the definition (A2.7)
 of $Z^{p.q}_r$).

 One can show that definition of $d_r$ is correct, $d_r^2=0$ and
 (A2.13) is obeyed [\Postn].

Using (A2.13) one come after finite number of steps to
    $E^{p.q}_\infty$ calculating each $E^{p.q}_r$  as
the cohomology group of the $E^{p.q}_{r-1}$:
 $E^{p.q}_1=H(d_0,E^{p.q})$,
    $E^{p.q}_2=H(d_1,E^{p.q}_1$ and so on.

    The spaces $E^{p.q}_r$  can be considered intuitively as
$r$--th order (with respect to differential $\p_2$)
 cohomologies of differential $Q$ .
The operator $\p_1$ is zeroth order approximation for
differential $Q$.
  The calculations of
 $E^{p.q}_\infty$ via (A2.13) can be considered as
perturbational calculations.
\smallskip
One can develop this scheme considering in perturbative
calculations not the operator $\p_1$, but $\p_2$ as
zeroth order approximation.

Instead filtration (A2.4) one has consider the
"transposed" filtration
                          $$
    \dots\subseteq\, ^t X^m\subseteq\,
  ^t X^{m+1}\subseteq\dots\subseteq\, ^t X^1\subseteq X^0
                                              \eqno (A2.14)
                           $$
                     $$
   {\rm where}\qquad\quad
 \,^t X^k= \bigoplus_{p\geq 0,q\geq k} E^{p.q}
                    $$
   and corresponding transposed
  spaces $\{\,^t E^{p.q}_r\}$. For example
                      $$
                  E^{p.q}_1=H(\p_1, E^{p.q}),\quad
                   ^t E^{p.q}_r=H(\p_2, E^{p.q})\,.
                       $$
Instead spectral sequence $\{E^{*.*}_r,d_r\}$ one has to consider
 transposed spectral sequence $\{^tE^{*.*}_r,\,^td_r\}$:
                             $$
               d_0=\p_1,\rightarrow \,^td_0=\p_2\,;
               d_1[c]_1=[\p_2 c]_1,\rightarrow \,
              ^td_1 [c]_1= [\p_1 c]_1\,,
                         $$
and so on.

  The relations between spaces
 $\{E^{p.q}_\infty\}$ and $\{^t E^{p.q}_\infty\}$
 which express in different ways
the cohomology $H(Q)$ is one of the applications of the
method described here.

 {\bf Example.} Let ${\bf c}=(c_0,c_1.c_2)$ where
 $c_0\in E^{0.2}, c_1\in E^{1.1}, c_2\in E^{2.0}$ be cocycle
 of the differential $Q$:
$Q(c_0,c_1.c_2)=0$ i.e. $\p_1 c_0=0,\p_2 c_0=-\p_1 c_1,
                        \p_2 c_1=\p_1 c_2$.
 To the leading term $c_0$ of this cocycle w.r.t. the
filtration (A2.4) corresponds the element
 $[c_0]_\infty$ in $E^{0.2}_\infty$
which  represents the cohomology class
 of the cocycle ${\bf c}$ in
   $E^{0.2}_\infty$.

In the case if the equation
$(c_0,c_1.c_2)+Q(b_0,b_1)=(0,c_1^\prime,c_2^\prime)$
 has a solution, i.e. the leading term $c_0$
 of the cocycle ${\bf c}$  can be cancelled
  by changing of this cocycle on a coboundary, then
the element   $[c_1^\prime]_\infty\in E^{1.1}_\infty$
 represents the cohomology class of the cocycle ${\bf c}$ in
   $E^{1.1}_\infty$.

 In the case if the equation
$(c_0,c_1.c_2)+Q(b_0,b_1)=(0,0,\t c_2)$
have a solution, i.e.
the leading term and next one
both can be cancelled, by redefinition
on a coboundary, then
  $[{\t c_2}]_\infty\in E^{2.0}_\infty$
 represents the cohomology class of the cocycle ${\bf c}$ in
   $E^{2.0}_\infty$.

  To put correspondences between
  the cohomology class of the cocycle ${\bf c}$
 and corresponding elements from
 transposed spaces
$^tE^{0.2}_\infty,\,^t E^{1.1}_\infty\,^tE^{1.1}_\infty$ we
have to do the same, changing only the definition of leading terms,
which we have to consider now w.r.t. the filtration (A2.14).

 To the leading term $c_2$ of this cocycle w.r.t. the
filtration (A2.14) corresponds the element
 $[c_2]_\infty$ in $\,^tE^{2.0}_\infty$
which  represents the cohomology class
 of the cocycle ${\bf c}$ in
   $\,^tE^{2.0}_\infty$.
In the case if the equation
$(c_0,c_1.c_2)+Q(b_0,b_1)=(c_0^\prime,c_1^\prime,0)$
 has a solution, i.e. the leading term $c_0$
 of the cocycle ${\bf c}$  can be cancelled
  by changing of on a coboundary, then
the element $[c_1^\prime]_\infty$
 represents the cohomology class of the cocycle ${\bf c}$ in
   $\,^tE^{1.1}_\infty$.
 In the case if the equation
 $(c_0,c_1.c_2)+Q(b_0,b_1)=(\t c_0,0,0)$
 has a solution, then $[{\t c_0}]$
 represents the cohomology class of the cocycle ${\bf c}$ in
   $\,^tE^{0.2}_\infty$.
   \bigskip

       \centerline {\bf References}

 [\Fad]  L.D.Faddeev, S.L.Shatashvili,
 "Algebraic and Hamiltonian methods
 in the Theory of non-Abelian Anomalies",
         Theor.Math.Phys.,
               {\bf 60}, 206-217, (1984).
\medskip
[\Jack] R.Jackiw, "Anomalies and Cocycles",
 Commun. Nucl. Part. Phys. {\bf 15}, 99--116, (1985).
  \medskip
[\Kost] B.Kostant, S.Sternberg, "Symplectic reduction, B.R.S.
 cohomology and infinite--di\-men\-si\-onal Clifford algebras"
  Annals of Physics, {\bf 176}, 49--113, (1987).
  \medskip
[\Dub] M.Dubois--Violette, "Systems dynamiques contraints: L'approche
 homologique"
  Ann. Inst. Fourier, {\bf 37},4, pp.45--57, (1987).
 \medskip
 [\Mcmull] A.D.Browning, Mc--Mullan,D.
 "The Batalin Fradkin amd Vilkovisky
 Formalism for higher order theories",
   J.Math.Phys. {\bf 28 }, 438 (1987)
 \medskip
[\Hen] M. Henneaux, C.Teitelbom,  "BRST Cohomology in Classical
 Mechanics", Commun. Math. Phys., {\bf 115}, 213--230, (1988).
  \medskip
[\Stash]  J.Fish, M.Henneaux, J.Stasheff, C.Teitelbom,
  "Existence, Uniqueness and Cohomology of the Classical BRST Charge
 with Ghosts of Ghosts",
 Commun. Math. Phys., {\bf 120}, 379--407 (1989).
 \medskip

[\Dix] J.A.Dixon,
  "Calculation of BRS cohomology with Spectral Sequences",

 \noindent Comm.Math.Phys., {\bf 139}, 495--525, (1991).
   (Cohomology and Renormalization of Gauge Theories, I, II, III,
   unpublished preprints, 1976--1977.)
\medskip
[\Barn] G.Barnich, F.Brandt, M.Henneaux,
   "Local BRST Cohomology in the Antifield Formalism",
 \noindent Comm.Math.Phys., {\bf 174}, 57--93, (1995).
\medskip
[\Pig] O.Piguet, S.P.Sorella,
 "Algebraic Renormalization", Lecture Notes in Physics.

\noindent Springer-Verlag 1995.

 \medskip
 [\Arn] V.I.Arnold,  "Mathematical methods of classical mechanics"--
 Moscow, Nauka (1974).
  \medskip
 [\Ann] A.Cabo, J.L.Lucio., M. Napsuciale,
 "Cocycle structure and Symmetry Breaking
 in Supersymmetric Quantum Mechanics"
 Annals of Physics {\bf 244}, 1--11, (1995).
   \medskip

[\Vor] T.Voronov,
 "The Complex Generated by Variational Derivatives,
  Lagrangian Formalism of
 Infinite Order and Stokes Formula Generalization"

 Uspekhi Mat. Nauk (in Russian) , n.6, 195--196, (1996).
  \medskip
 [\Kh]  A.V.Gayduk, O.M.Khudaverdian, A.S.Schwarz,
 "Integration over Surfaces in a \break Superspace",
 Theor.Math.Phys. {\bf 52}, 375--383, (1982).

 \medskip
[\Voron] T.Voronov,  "Geometric Integration Theory on supermanifolds"
   Sov.Sci.Rev.C Math {\bf 9}, 1--138, (1992).

  \medskip
[\Olv]  P.J. Olver, "Applications of the Groups to
 Differential Equations" Springer--Verlag, (1986).
  \medskip
[\Khud]  O.M.Khudaverdian, A.P.Nersessian,
 "Batalin--Vilkovisky Formalism and
  Integration Theory on a Manifolds",
  J. Math. Phys. {\bf 37}, 3713--3724, (1996).
       \medskip
 [\Sch] A.S. Schwarz, "Are the Field and Space Variables
  on an Equal Footing?",

 \noindent  Nucl.Phys., {\bf B171}, 154--166, (1980).
   \medskip
 [\Stern] V. Guillemin, S. Sternberg, "Geometric asymptotics"
 Math. surv. N.14, 1977, AMS, Providence, Rhode Island.
\medskip
 [\Ners]  A.Nersessian, V.M.Ter-Antonyan,
 "Anyons, Monopole and Coulomb Problem", physics/9712027.
\medskip
[\Stor]  C. Becchi, A. Ruet, R.Stora,
 "Renormalizable Theories with Symmetry Breaking",
in: "Field Theory, Quantization and Statistical Physics"
 (ed. E. Tirapegui) D. Reidel, Dordrecht, 1981.
  \medskip
 [\Vin]  "Symmetries and Conservations Laws for
Mathematical Physics Equations". Editing
by A.M. Vinogradov, I.S.Krasils`shik.
  Moscow, 1997.
\medskip
 [\Postn]  M.M.Postnikov, "Lectures on Geometry, III, V"
  Œoscow, Nauka, (1987).
\medskip
[\Sta] J.Stasheff  "The (secret) Homological algebra of the Batalin-
 Vilkovisky Formalism"  Proceedings of the Conference
   {\it Secondary Calculus and Cohomological Physics},
   Moscow, 1997.
  \bye